\documentclass[twocolumn,showpacs,preprintnumbers,amsmath,amssymb]{revtex4}

\usepackage{graphicx}% Include figure files
\usepackage{dcolumn}% Align table columns on decimal point
\usepackage{bm}% bold math
\usepackage{epsfig}
\usepackage{epstopdf}
\usepackage{grffile}
\usepackage{color}
\usepackage{colordvi}
\usepackage{amsmath,amssymb}
\usepackage{rotating}
\usepackage{lscape}

\usepackage{hyperref}

\newcommand{\be}[0]{\begin{equation}}
\newcommand{\ee}[0]{\end{equation}}
\newcommand{\ba}[0]{\begin{eqnarray}}
\newcommand{\ea}[0]{\end{eqnarray}}

\begin{document}

%
%=====================================================================================
%
\title{ New parton distributions in fixed flavour factorization
        scheme from recent deep-inelastic-scattering data }
\author{H. Khanpour$^{a,b}$ }
\email{Hamzeh.Khanpour@nit.ac.ir }
\author{Ali N. Khorramian$^{a,b}$ }
\email{Khorramiana@theory.ipm.ac.ir }
\author{S. Atashbar Tehrani$^{b}$ }
\email{Atashbar@ipm.ir}

\affiliation{
$^{(a)}$Physics Department, Semnan University, Semnan, Iran  \\
$^{(b)}$School of Particles and Accelerators, Institute for
Research in Fundamental Sciences (IPM), P.O.Box 19395-5531,
Tehran, Iran}

\homepage{http://particles.ipm.ir/ }

\date{\today}

\begin{abstract}

We present our QCD analysis of the proton structure function
$F_2^p(x,Q^2)$ to determine the parton distributions at the next-to-leading order (NLO).
The heavy quark contributions to $F_2^i(x,Q^2)$, with $i$ = $c$, $b$  have been included in
the framework of the `fixed flavour number scheme' (FFNS).
The results obtained in the FFNS are compared with available results such as the general-mass
variable-flavour-number scheme (GM-VFNS) and other
techniques used in global QCD fits of parton distribution functions. In the present QCD analysis, we use a wide range of
the inclusive neutral-current deep-inelastic-scattering (NC DIS) data, including the most recent
data for charm $F_2^c$, bottom $F_2^b$, longitudinal $F_L$ structure functions and also the
reduced DIS cross sections $\sigma_{r,NC}^\pm$ from HERA experiments.
The most recent HERMES data for proton and deuteron structure functions are also added.
We take into account ZEUS neutral current $e^ \pm p$ DIS inclusive
jet cross section data from HERA together with the recent Tevatron Run-II inclusive jet cross section data from CDF and D{\O}.
The impact of these recent DIS data on the PDFs extracted from the global fits are studied.
We present two families of PDFs, {\tt KKT12} and {\tt KKT12C}, without and with
HERA `combined' data sets on $e^{\pm}p$ DIS. We find these are in good agreement with the available theoretical models.

\end{abstract}



\maketitle

%\tableofcontents{}

%%%%%%%%%%%%%%%%%%%%%%%%%%%%%%%%%%%%%%%%%%    Introduction    %%%%%%%%%%%%%%%%%%%%%%%%%%%%%%%%%%%%%%%%%%%%%%%%%
%
\section{Introduction} \label{Introduction}
Lepton nucleon deep inelastic scattering (DIS) processes play a very important role for our understanding of quantum
chromodynamics (QCD) and nucleon structure \cite{DeRoeck:2011na}.
The DIS process of leptons off nucleons is an
instrumental tool for high precision measurements of the quark and
gluon content of the nucleons.
DIS experiments provide valuable information from lepton-nucleon scattering cross
sections to find out about the structure functions of targets.
Note that these structure functions are related to parton distribution functions (PDFs) directly.
This process at large momentum transfer provides one of the cleanest possibilities to test the
predictions of QCD and allows us to measure the high-precision PDFs
of the nucleons together with the strong coupling constant.  DIS process also
demonstrates the internal structure of the nucleon in high-energy
scattering at the CERN Large Hadron Collider, Fermilab Tevatron
collider, and in other hard scattering experiments. The advent and development
of the electron-proton colliders, especially HERA, has resulted in
significant progress in the recent years on the precise
understanding of the structure of the nucleon, especially in different
kinematic regions of momentum fractions $x$ of proton carried by the struck parton.
The most precise inclusive structure function measurements which has been provided
by HERA collider, has brought remarkable improvements to our understanding
on proton structure and can be used to extract high precision
PDFs, which is important for any process which involves colliding hadrons.

Recently it is quite common for several  theoretical groups
to make a global fit on available experimental data to extract parton distribution functions which
are utterly useful in colliding hadrons, specially LHC, studies. Nowadays thousands of
published  data points from various experiments are available to extract more precise PDFs
and strong coupling constants in higher perturbative orders.
Also for more accuracy, the QCD fits can be performed with leading-order (LO), next-to-leading
order (NLO) and even next-to-next-to-leading order (NNLO) QCD approximation.
In addition, it is also significant to find out about PDFs uncertainties for experimentalist.

In recent years, a number of theoretical groups providing global QCD fits (CTEQ/CT10~\cite{Tung:2006tb,Lai:2010vv} and MRST/MSTW08~\cite{Martin:2009bu,Martin:2010db,Martin:2009iq})
have developed a prescription for determining uncertainties, and
have now been joined by several other theoretical groups (NNPDF~\cite{Ball:2012cx,Ball:2010de,Ball:2011mu}, ABKM10~\cite{Alekhin:2009ni}, GJR08/JR09~\cite{Gluck:2007ck,JimenezDelgado:2008hf}).
HERA experimental collaborations have combined their DIS data and published parton
distribution functions, (HERAPDF), with uncertainty error bands as well~\cite{:2009wt,CooperSarkar:2010ik}.

Nowadays, new PDFs incessantly become more accurate by considering
up-to-date theoretical developments and  data from hadron colliders experiments, also such evolvement
provide more reliable calculation of uncertainties corresponding to different theoretical and experimental inputs.
The correct treatment of heavy quark flavours in global QCD analysis of
PDFs is essential for precision measurements at hadron colliders such as HERA,
since it contributes up to 30\% to the DIS inclusive cross section
at small $x$.
This is very important for the reliability of parton distributions in the kinematic region
corresponding to the foreseen experiments at LHC since DIS remains an incomparable source of
information about parton distributions at small $x$, despite the effect of collider data on the PDFs.
Recent reported results \cite{Alekhin:2012du,Botje:2011sn,Alekhin:2011sk,Martin:2010db,cteq65,MSTW,cteq66,Watt08} show that the cross
section of Higgs, W and Z production at the LHC are sensitive to detailed aspects of PDFs that
depend on the heavy quark mass effects; and standard model processes as well as beyond standard models considerably
depend on improved knowledge of the c-quark parton density, in addition to the light flavours density \cite{Thorne:2008xf}.
Heavy quark flavours production in DIS is calculable in QCD and provides significant information on the gluonic content
of the proton. Nowadays, heavy quarks are produced in several
experiments of high energy physics and the production and decay
properties of the heavy quarks are extremely interesting.
A number of theoretical groups, such as CTEQ collaboration, CT10
\cite{Lai:2010vv}, MSTW08 \cite{Martin:2009iq}, ABKM10
\cite{Alekhin:2009ni} and GJR08 \cite{Gluck:2007ck} have produced
publicly available PDFs using different data sets.
We also studied the diffractive process to parameterize the diffractive parton distribution
functions (DPDFs) taking into account heavy quarks contributions using recent experimental data \cite{Monfared:2011xf}.

In the present paper we will extract the PDFs using QCD analysis of
a large number of available and up-to-date
experimental data in the frame work of FFNS by introducing
a new kind of parametrization.
This paper deals with the construction of a FFNS and
the comparison of their predictions with the
most recent experimental data for various region of $x$ and $Q^2$.
Since we have recent experimental data for the reduced cross section, heavy quarks and longitudinal
structure functions from the HERA collaboration, and new fixed target data for
proton and deuteron structure functions from HERMES, we have enough motivation
to extract all of parton distribution functions using
new sets of parametrization with take into account the heavy quark
contributions.

The organization of the paper is as follows. In Sec.~\ref{theoretical-framework},  we
outline the theoretical formalism which describes the DIS
structure functions. Our parametrization and heavy quark contributions to the proton
structure function and the formulation of FFN scheme are also
performed in this Section. In Sec.~\ref{parton-analyses}, we discuss the global
parton analysis and then present the results of our NLO PDF fit to
the DIS world data to determine the PDF
parameters and $\alpha_{s}(M_{Z}^{2})$. Discussion of fit results
are given in Sec.~\ref{fit-results}.
This section also contains a detailed comparison of
the {\tt KKT12} and {\tt KKT12C} QCD fits with other recent PDF sets and with the recent HERA data in various $x$ and $Q^2$ regions.
The results related to the longitudinal
structure function $F_L(x,Q^2)$ is presented in Sec.~\ref{longitudinalSF}. Section~
\ref{Summary} contains our summary and conclusions.

%
%%%%%%%%%%%%%%%%%    Theoretical background of the QCD analysis   %%%%%%%%%%%%%%%%%%%%%
%
\section{Theoretical background of the QCD analysis} \label{theoretical-framework}
In this section, we first give a brief overview of the standard
theoretical formalism that we used in the present QCD analysis. A more detailed description is given
later in separate sections.

\subsection{Theoretical input} \label{Theoretical-input}
The QCD predictions for the structure functions are obtained by solving the well known DGLAP evolution
equations at NLO in the modified minimal
subtraction ($\overline{\rm MS}$) scheme. We choose the factorization and renormalization
scales to be $Q^2$. We work in the $\overline{\rm MS}$ scheme, where the
deep-inelastic nucleon structure function
$F_2(x,Q^2)$, can be written as a convolution of the parton distribution functions
together the coefficient functions \cite{Martin:2009iq}.
The scale dependence of the PDFs is given by the  well-known DGLAP evolution
equations which yield the PDFs at all values of $Q^2$ if they are provided as functions
of $x$ at the input scale of $Q_0^2$.

A number of methods to solve the evolution equations have been
proposed, including direct $x$-space method, Mellin-transform
methods and orthogonal polynomials methods. More detailed about
the later method are given in Refs.
\cite{Khorramian:2009xz,Khanpour:2009zz,Khorramian:2008yh,Khorramian:2007zz,
Khorramian:2008zz,Khorramian:2008zza,AtashbarTehrani:2009zz,
AtashbarTehrani:2007be,Mirjalili:2007ep,Mirjalili:2006hf,Khanpour:2011zz,Khanpour:2010zz}.

In the present paper we have performed all Q$^2$-evolutions in Mellin $N$-moment space for the NLO evolution.
In this regard, the program QCD-PEGASUS \cite{Vogt:2004ns} is used in order to perform all
Q$^2$-evolutions of parton distributions.
The input for the evolution equations and the strong coupling constant $\alpha_s(Q_0^2)$, at a reference input scale
taken to be Q$_0^2$ = 2 GeV$^2$, must be extracted from a global QCD analysis of DIS world data.

Our present standard NLO QCD analysis has been performed within the $\overline{\rm MS}$ factorization and
renormalization scheme. We do not consider the heavy quarks ($c$, $b$, $t$) as massless partons within the nucleon.
In this case, the number of active (light) flavours $n_f$ appearing in the Wilson coefficients and the corresponding splitting
functions to be fix at $n_f = 3$. Furthermore, the evaluation of the running strong
coupling $\alpha_s(Q^2)$ needs to match $\alpha_s^{(n_f)}$ at
$Q= m_h  \, \,  (m_c = 1.41\,{\rm GeV}, m_b = 4.50\,{\rm GeV}, m_t
= 175\,{\rm GeV})$,  i.e. $\alpha_s^{(n_f)} (m_h) = \alpha_s^{(n_f-1)} (m_h)$.

%
%%%%%%%%%%%%%%%%%%%%%%%%%%%%%%%%%%%%%%%%%%    Parameterisation    %%%%%%%%%%%%%%%%%%%%%%%%%%%%%%%%%%%%%%%%%%%%%%%%%
%
\subsection{Parametrization} \label{Parametrization}
For the QCD fit in the present analysis we use the following
independent functional form of the
parametrization of the parton distribution functions at the input scale
$Q_0^2$=2 GeV$^2$ for the valence quark distributions $xu_v$ and
$xd_v$, the anti-quark distributions $xS = 2x(\bar{u}+\bar{d}+\bar{s})$, $x\Delta = x (\bar{d} - \bar{u}
)$, and for gluon distribution $xg$:
\begin{eqnarray} \label{PDFs}
xu_v(x,Q_0^2)    &=& A_u\, x^{\alpha_u}(1-x)^{\beta_u}(1+\gamma_u\, x^{\delta_{u}} + \eta_u\, x),  \nonumber \\
xd_v(x,Q_0^2)    &=& A_d\, x^{\alpha_d}(1-x)^{\beta_d}(1+\gamma_d\, x^{\delta_{d}} + \eta_d\, x),  \nonumber \\
x\Delta(x,Q_0^2) &=& A_{\Delta}\, x^{\alpha_{\Delta}}(1-x)^{\beta_S + \beta_{\Delta}}
(1+\gamma_{\Delta}\, x^{\delta_{\Delta}} + \eta_{\Delta}\, x),  \nonumber \\
xS(x,Q_0^2)      &=& A_S\, x^{\alpha_S}(1-x)^{\beta_S}(1+\gamma_S\, x^{\delta_S} + \eta_S\, x),    \nonumber \\
xg(x,Q_0^2)      &=& A_g\, x^{\alpha_g}(1-x)^{\beta_g}(1 +\gamma_g\, x^{\delta_g} + \eta_g\, x ).
\end{eqnarray}
The flavour structure of the light quark sea at $Q_0^2$ is taken
to be
\begin{eqnarray}\label{PDFsNEWc}
%\begin{array}
2 \bar{u}(x,Q_0^2) &\, =\, & 0.4 S(x,Q_0^2) - \Delta(x,Q_0^2), \\
2 \bar{d}(x,Q_0^2) &\, =\, & 0.4 S(x,Q_0^2) + \Delta(x,Q_0^2), \\
2 \bar{s}(x,Q_0^2) &\, =\, & 0.2 S(x,Q_0^2),
%\end{array}
\end{eqnarray}
and we take a symmetric strange sea $s(x,Q_0^2) = \bar s(x,Q_0^2)$ due to
the insensitivity of data sets we are using to the specific choice of the strange quark distributions.
Due to the very large amount of data in our global QCD fits and the presence of some constrains,
we can control the parameters in order to allow
sufficient flexibility in the form of the parton distributions.

In the present work, the input PDFs listed in Eq.~\ref{PDFs} are subject to three constraints from
valence quarks number sum rules together with the total momentum sum rule \cite{Martin:2009iq}.
Not all the parameters in our inputs for the parton distributions
are free. The parameters $A_u$, $A_d$ and $A_g$ were calculated
from the other parameters using mentioned constrains, therefore there
are potentially 28 free parameters in the PDF fit, including $\alpha_s(Q_0^2)$.

The mentioned parametrization for the input PDFs in our analysis,
Eq.~\ref{PDFs}, and all free PDF parameters listed there allow the fits a large degree of flexibility.
In the final minimization, we fixed some parameters and only 13
parameters remained free for all parton flavour in the fits. These
parameters are allowed to go free in order to calculate the covariance matrix.

According to our parameterization in  Eq.~\ref{PDFs} and for valence quark distributions, $xu_v$ and $xd_v$,
we use the standard functional form. The normalization parameter $A_v$ is obtained from the number
of valence quarks sum rules. The variation of the valence distributions at intermediate values of $x$ will be
controlled by the third free parameter $\gamma_v$. In our calculation for valence sector, we found that the
parameter $\eta_v$ is correlated to $\alpha_v$ and $\beta_v$ more strongly.
At very high and even at very small $x$, the uncertainties are constrained by the mentioned sum rules. In the region of
$x$ =[0.01, 0.75], where the valence PDFs are constrained by DIS data, about 75\% of the total number of valence quarks can be found.
We should notice that in the reported results of MSTW08 \cite{Martin:2009iq} and GJR08 \cite{Gluck:2007ck} for input parametrization, the parameter
value of $\delta_{v}$ is fixed to 0.5 which we want to deal with
this parameter as a free parameter. After first minimization, in both cases of valence quark
distributions, we will get the parameter values of $\gamma_v$, $\delta_{v}$ and $\eta_v$ fixed
so only two parameters remain free, $\alpha_v$ and $\beta_v$.

In order to consider the flavour separation of the sea-quark
distributions $x\Delta = x (\bar{d} - \bar{u} )$
in Eq.~\ref{PDFs}, we choose $\beta_S + \beta_{\Delta}$ as a power of
$(1-x)$ for  $x\Delta$, because at high $x$ this
PDF is becoming very small and we attempt to constrain $\bar{u}$
and $\bar{d}$ to be positive. We need also to have $x^{\delta_{\Delta}}$ term instead  of
the more usual $x^{0.5}$ and $x^{2}$ term to give sufficient
flexibility at medium and large $x$. In fact we consider
$A_{\Delta}$, $\alpha_{\Delta}$ and
$\beta_{\Delta}$ as free parameters and $\gamma_{\Delta}$, ${\delta_{\Delta}}$ and
$\eta_{\Delta}$ will be fixed at their best-fit values.

In the sea quark distribution, $xS = 2x(\bar{u}+\bar{d}+\bar{s})$ in Eq.~\ref{PDFs},
there is no sum rule constraint, so in principle there are
3 free parameters including  $A_S$, $\alpha_S$ and
$\beta_S$ and the other 3 parameters will expected
to be fixed since the data do not constrain these
parameters well enough.

The gluon distribution is a far more difficult case for PDFs
parameterizations to obtain precise information.
Since the current data provide little constraint on the gluon density,
we expect the functional form of gluon distributions
in Eq.~\ref{PDFs} to allow us for better gluon PDF determination.
The input gluon in the MSTW08 \cite{Martin:2009iq} is almost the same form
of the original MRST fits \cite{Martin:1999ww,Martin:1998sq}, but MSTW08 added a
second term to avoid a negative gluon distribution at very
small $x$. We found that the parameter $\delta_{g}$
in our gluon parametrization could help us
to facilitate this. In this case we need to have a
$x^{{\delta_{g}}=2}$ term instead of the more usual $\sqrt{x}$
term to give sufficient flexibility at low $x$, and we find stability in the QCD fits.
On the other hand, one feature of our parametrization is that we allow
the gluon to have a very general form over a wide range of $x$,
specially at medium $x$, because it can induce potentially large
effects on the strong coupling constant obtained from the global
fits and indirectly leads to a change in the light partons.
In particular, fixing the parameters $\gamma_g$,
$\delta_{g}$ and $\eta_g$ at the best fit values we
find that $\alpha_g$ and $\beta_g$ are sufficient for the gluon -- one for high $x$ and one for low $x$.
Consequently in the final minimization, only two parameters $\alpha_g$ and $\beta_g$ remain free. We note that
we can determine $A_g$  by the total momentum sum rule \cite{Martin:2009iq}.

As a last point, we note that the single-inclusive jet data are very important for the constrains of gluon distribution.
The constraints on the gluon distributions in the domain $ 0.01 \lesssim x \lesssim 0.5 $ are provided by
the inclusive jet production data from Run II at the Tevatron from CDF and D{\O}
while the data from HERA on inclusive jet production in DIS constrain the gluon for $ 0.01 \lesssim x \lesssim 0.1 $.

%
%%%%%%%%%%%%%%%%%%%%%%%%%%%%%%%%%%%%%%%%%%    Treatment of heavy flavours    %%%%%%%%%%%%%%%%%%%%%%%%%%%%%%%%%%%%%%%%%%%%%%%%%
%
\subsection{Treatment of heavy flavours} \label{heavy-flavours}
In this section, we will discuss the fixed flavour number
(FFN) scheme  parton model predictions at high energy colliders at
LO and NLO of QCD. It is expected that perturbative QCD at NLO should give a good
description of heavy quark flavour production in hard scattering processes, so the correct
treatment of heavy flavours in common QCD analysis of PDFs at this
order is essential for precision measurements at hadron colliders.
The LO and NLO heavy quark contributions $F_i^{c,b} (x,Q^2)$ to the proton structure function are
calculated in the FFN approach and contribute to the total structure
functions as $F_i (x,Q^2)  = F_i^{\rm light}+ F_i^{\rm heavy}$
where the $F_i^{\rm light}$ refers to the common $u,\, d,\, s$ (anti) quarks and
gluon initiated contributions, and $F_i^{\rm heavy} = F_i^c
+F_i^b$ with $i = 2, L$. Charm and bottom quark contributions to
the total DIS cross section at high $Q^2$ are up to 30\% and 3\%,
respectively, and top quark contributions are negligible at present
energies. As already mentioned at the end of Sec.~\ref{Theoretical-input},
we employ for our analysis the FFN scheme
and fix the number of active light flavours $n_f$ = 3 in all
splitting functions $P^{(k)}_{ij}$ and in the corresponding Wilson
coefficients. In this factorization scheme only the light quarks
($u$, $d$, $s$) are genuine, i.e., massless partons within the
nucleon, whereas the heavy ones ($c$, $b$, $t$) are not.

Fixed flavour number approach is the region where the hard scale of
the DIS process is around the quark mass, i.e.~$Q^2\lesssim m_h^2$
where $h = c, b, t$.  In this case, massive quarks are not considered as partons within the proton.
They are described as final-state particles and so they
are entirely created in the final-state. Our practical reason for
extracting partons from the fits to structure function data in the
fixed three-light-flavour ($n_f = 3$) is to have a set of
up-to-date partons in order to make consistent comparisons between
other parton sets available using the same theoretical framework
of other scheme such as general mass variable-flavour number
scheme (GM-VFNS) \cite{Lai:2010vv,Martin:2009iq} and FFNS
\cite{Gluck:2007ck}. In addition for many exclusive or
semi-inclusive processes the theoretical predictions for the hard
scattering cross sections are only available in the FFN scheme.
The different heavy quarks treatment which nearly every group does,
may led to different results for the extracted parton
distribution functions.

%
%%%%%%%%%%%%%%%%%%%%%%%%%%%%%%%%%%%%%%%%%%    Fixed flavour number scheme ..........  %%%%%%%%%%%%%%%%%%%%%%%%%%%%%%%%%%%%%%%%%%%%%%%%%
%
\subsection{Fixed flavour number scheme (FFNS) and $F_2$ structure function} \label{F-F-N-S}
The NLO QCD analysis presented in this paper is performed in the modified minimal subtraction
($\overline{\rm MS}$) factorization and renormalization scheme.
Heavy quarks ($c$, $b$, $t$) are not considered as massless partons within the nucleon and
all their effects are contained in the perturbative coefficient functions.
This defines the so-called ``fixed flavour number scheme" (FFNS),
which is fully predictive in the heavy quark sector and incorporates the correct threshold behavior.
In the common $\overline{\rm MS}$ factorization scheme, the relevant total
proton structure function $F_2^p (x,Q^2)$ as extracted from the DIS $ep$ process
can be, up to NLO, written as
\cite{Martin:2009iq,Alekhin:2009ni,Gluck:2007ck,JimenezDelgado:2008hf,Gluck:1998xa,Pisano:2008ms,Gluck:2006pm}
\begin{eqnarray} \label{eq1}
F_2(x,Q^2) &=& F_{2,{\rm NS}}^+(x,Q^2) + F_{2,S}(x,Q^2) \nonumber \\
 &+& F_2^{(c,b)}(x,Q^2,m_{c,b}^2)\,,
\end{eqnarray}
with the non--singlet contribution for three active (light) flavours are given by
\begin{eqnarray}
\frac{1}{x}\, F_{2,{\rm NS}}^+(x,Q^2) & = &
\Big[C_{2,q}^{(0)} + a_s C_{2,{\rm NS}}^{(1)} \Big]  \nonumber \\
&& \otimes \left[\frac{1}{18}\, q_8^+ +\frac{1}{6}\, q_3^+\right](x,Q^2)\,,
\end{eqnarray}
where $C_{2,q}^{(0)}(z)=\delta(1-z)$ and $C_{2,{\rm NS}}^{(1)}$ is the
common NLO coefficient function
\cite{vanNeerven:2000uj,vanNeerven:1999ca,Vermaseren:2005qc}\,.
The NLO $Q^2$-evolution of the flavour non--singlet
combinations $q_8^+=u+\bar{u}+d+\bar{d}-2(s+\bar{s}) =
u_v+d_v+2\bar{u}+2\bar{d}-4\bar{s}$  and $q_3^+=u+\bar{u}-(d+\bar{d})=u_v-d_v$
is related to the LO (1-loop) and NLO (2-loop)
splitting functions $P_{\rm NS}^{(0)}$ and $P_{\rm NS}^{(1)+}$ respectively
\cite{vanNeerven:2000uj,vanNeerven:1999ca}\,.

The flavour singlet contribution in Eq.~\ref{eq1} reads
\begin{eqnarray}
\frac{1}{x}\, F_{2,S}(x,Q^2)&=&\frac{2}{9} \left\{ \left[ C_{2,q}^{(0)} + a_s C_{2,q}^{(1)} \right]
\otimes \Sigma \right. \nonumber \\ &&
\left.   +  \, a_s C_{2,g}^{(1)} \otimes g\right\} (x,Q^2) \,,
\end{eqnarray}
with $\Sigma(x,Q^2)\equiv\Sigma_{q=u,d,s}(q+\bar{q})=u_v+d_v+2\bar{u}+2\bar{d}+2\bar{s}$,
$C_{2,q}^{(1)}=C_{2,\rm NS}^{(1)}$ and the additional NLO
gluonic coefficient function $C_{2,g}^{(1)}$ can be found in Refs.
\cite{vanNeerven:2000uj,Vermaseren:2005qc}\, for example.
Notice that we consider sea breaking effects ($\bar{u}\neq\bar{d}$) since the HERA data is used, and thus
our analysis are sensitive to these corrections. Since these data
sets are also insensitive to the specific
choice of the strange quark distributions, we consider a symmetric
strange sea, $s=\bar{s}$. We also use
$s = \bar{s} = \frac{\kappa}{2}\big( \bar{u}  + \bar{d} \big)$
where in practice $\kappa$ is a constant fixed to $\kappa = 0.4 - 0.5$.
The heavy flavour contributions $F_{2}^{c,b} (x,Q^2)$ that have been added
in the present analysis will be discussed in the next section.

%
%%%%%%%%%%%%%%%%%%%%%%%%%%%%%%%%%%%%%%%%%%    Heavy flavour contributions  %%%%%%%%%%%%%%%%%%%%%%%%%%%%%%%%%%%%%%%%%%%%%%%%%
%
\subsection{Heavy flavour contributions} \label{Heavy-flavour-contributions}

The heavy flavour contributions $F_{2} ^{c,b} (x,Q^2)$ which have been added in the present analysis are taken
as in Refs.~\cite{Gluck:2006pm,Gluck:2004fi,Gluck:2008gs,Gluck:1993dpa,Laenen:1992cc,Riemersma:1994hv,Laenen:1992zk}\,.
The structure functions in fixed flavour factorization scheme are written by
\begin{equation} \label{FFNS}
  F^h_2(x,Q^2) = \sum_i C^{{\rm FFNS}, n_f}_{2,i}(Q^2/m_h^2)\otimes f^{n_f}_i(Q^2) \, ,
\end{equation}
where $f_i^{n_f}(Q^2)$ represents only to the light partons ($i$ = $u$, $d$, ...),
$m_{h}$ refers to the mass of the heavy quarks, with $h$ = $c$ , $b$ and
$n_f$ denotes the number of active flavour.

The FFNS approach which we are using  in the present analysis, contains all the heavy quark mass dependent contributions.
In leading order of QCD, ${\cal{O}}(\alpha_s)$, the  well known FFN
scheme heavy quark flavour contributions $C^{{\rm FFNS}, n_f}_{2,i}(Q^2/m_h^2)$ in $ep$ collisions are dominated by the
photon gluon fusion process (BGF), \mbox{$\gamma^*g\to h\bar{h}$}
~ \cite{Leveille:1978px, Witten:1975bh,
Shifman:1977yb,Gluck:1979aw,Gluck:1994uf}. The FFNS heavy flavour
coefficient functions for neutral-current structure functions
containing $\ln (Q^2/m^{2}_h)$ terms are well known up to NLO.
Due to the complexity of the heavy flavour coefficients function caused by the non-zero mass of heavy quarks, $m_h$,
there are no published complete analytic expressions for all
the coefficient functions and the results are available in table form at this order \cite{Riemersma:1994hv}.
In particular we use the code provided in
Ref.~\cite{Riemersma:1994hv}, which combines known analytic
expressions together with grids for the more complicated
coefficient functions; they are represented in the $\overline{\rm
MS}$ factorization scheme. Up to ${\cal{O}}(\alpha^2_s)$, the
relevant heavy quark contributions to DIS structure functions $F^{\rm heavy}_{2,L}$ are
given by \cite{Riemersma:1994hv,Laenen:1992zk}
\begin{eqnarray}\label{eq:FFNS}
F^h_{j=2,L}(x,Q^2,m^2) = \frac{Q^2 \alpha_s}{4\pi^2 m^2}
\int_x^{z_{\rm max}} \frac{dz}{z}  \Big[\,e_h^2
f_g(\frac{x}{z},\mu^2) c^{(0)}_{j,g} \,\Big]   \nonumber \\
+ \frac{Q^2 \alpha_s^2}{\pi m^2}
\int_x^{z_{\rm max}} \frac{dz}{z}  \Big[\,e_h^2 f_g(\frac{x}{z},\mu^2)
(c^{(1)}_{j,g} + \bar c^{(1)}_{j,g} \ln \frac{\mu^2}{m^2})  \, \, \,   \nonumber \\
+ \sum_{i=q,\bar q} \Big( e_h^2\,f_i(\frac{x}{z},\mu^2)
(c^{(1)}_{j,i} + \bar c^{(1)}_{j,i} \ln \frac{\mu^2}{m^2})   \hspace{1.8cm}  \nonumber \\
+ e^2_{L,i}\, f_i(\frac{x}{z},\mu^2) (d^{(1)}_{j,i} + \bar
d^{(1)}_{j,i} \ln\frac{\mu^2}{m^2}) \, \Big)  \,\Big] \,,  \,  \hspace{1.8cm}
\end{eqnarray}
where the upper boundary on the integration is given by $z_{\rm
max} = Q^2/(Q^2+4m^2)$ and sum runs over all the light quark and
antiquark flavours. Furthermore, $f_i(x,\mu^2)\, (i = q, \bar q)$
denote the light parton densities in the proton and $\mu$ stands for the
mass factorization scale, which normally has been put equal to the
renormalization scale. $e_{h}$ is the charge of the heavy quarks,
with $h = c , b$. The coefficient functions, represented by
$c^{(l)}_{j,i}(\eta, \xi)\,,\bar c^{(l)}_{j,i} (\eta, \xi)\,,(i =
g\,, q\,,\bar q \, ; l = 0, 1 \, ; j = 2, L )$ and by
$d^{(l)}_{j,i}(\eta, \xi)\,,\bar d^{(l)}_{j,i}(\eta, \xi)$, $(i =
q\,,\bar q\,;l = 0, 1 \, ; j = 2, L)$ are calculated in
\cite{Laenen:1992zk} and they are represented in the
$\overline{\rm MS}$ scheme. The LO ${\cal{O}}(\alpha_s)$
contributions to $F_{2,L}^h$, due to the subprocess $\gamma^*g\to
h\bar{h}$ with  $h=c,b$, have been summarized in
\cite{Gluck:1994uf}, and the NLO ${\cal{O}}(\alpha_s^2)$ ones are
given in \cite{Riemersma:1994hv,Laenen:1992zk}.

Different groups use various values for the charm and bottom quark masses varying
from 1.3 GeV to 1.65 GeV and from 4.3 GeV to 5 GeV, respectively.
This variation will change the extracted parton distribution functions.
A detailed discussion on mass dependence of parton distribution functions can be found in \cite{Martin:2010db,Ball:2011mu,Alekhin:2011jq,Alekhin:2010sv}.

In the present analysis we fix the heavy quark masses at $m_c = 1.41\,{\rm GeV}$ and $m_b = 4.50\,{\rm GeV}$,
this differs from the MRST and MSTW08 default values of $m_c = 1.43$ GeV, $m_b = 4.30$ GeV  and $m_c = 1.40$ GeV, $m_b = 4.75$ GeV, respectively.
The results of Ref.~\cite{Martin:2009iq} indicated that if we allow the mass of charm quark as a free parameter in the global fits,
one can find the best-fit value to be $m_c = 1.39$ GeV at NLO.

%
%%%%%%%%%%%%%%%%%%%%%%%%%%%%%%%%%%%%%%%%%%    Global parton analyses    %%%%%%%%%%%%%%%%%%%%%%%%%%%%%%%%%%%%%%%%%%%%%%%%%
%
\section{Global parton analyses} \label{parton-analyses}
%
%%%%%%%%%%%%%%%%%%%%%%%%%%%%%%%%%%%%%%%%%%    Choice of data sets    %%%%%%%%%%%%%%%%%%%%%%%%%%%%%%%%%%%%%%%%%%%%%%%%%
%
\subsection{Overview of data sets} \label{data-sets}
PDFs are usually determined in a global QCD analyses of a wide variety of DIS data,
both from HERA and fixed-target experiments as well as
$\nu(\bar{\nu})N$ $xF_3$ data from CHORUS, NuTeV and data for the longitudinal structure function, $F_L(x,Q^2)$.
In the recent years, several new precise data sets became available which
we include in our present global QCD fit.
In the present analysis we fit to the measured neutral current deep inelastic $e ^ \pm p$ scattering cross section
values that is defined by three structure functions, $F_2$, $F_L$ and $xF_3$ as \cite{DeRoeck:2011na,Gluck:2007ck}
\begin{equation} \label{Sigma}
 \sigma_{r, {\rm NC}} ^\pm (x,Q^2) = F_2^{\rm NC}(x,Q^2) - \frac{y^2}{Y_+}F_L^{\rm NC}(x,Q^2) \mp \frac{Y_-}{Y_+} {xF_3^{\rm NC}},
\end{equation}
where $0 < y < 1$ is the process inelasticity $ y = Q^2/(xs)$ and $s$ is the $e p$ centre-of-mass energy and $Y_\pm = 1 \pm (1-y^2)$.

%
%
%=====================================================================================
%
\begin{table*}
\centering
{\small
%{\footnotesize
{\begin{tabular}{|l|l|l|c|c|c|}
\hline  \hline
\textbf{Experiment}         &  \textbf{$x$--range}  & \textbf{$Q^{2}$--range \lbrack GeV$^{2}$]}&   \# of data points   &   \textbf{$\Delta {\cal N}_{n}$}   &   \textbf{${\cal N}_{n}$}   \\   \hline  \hline
%--------------------------------------------------------------------------------
%--------------------------------------------------------------------------------
H1 high  Q$^2$ 94--97 $e^+p$ NC    &0.003 -- 0.65    &150.0 -- 30000.0&   130   \cite{Adloff:1999ah}      &    1.5\%      & 0.9991        \\
H1 high  Q$^2$ 98--99 $e^-p$ NC    &0.0032 -- 0.65   &150.0 -- 30000.0&   126   \cite{Adloff:2000qj}      &    1.8\%      & 0.9999        \\
H1       Q$^2$ 96--97 $e^+p$ NC    &5.0E-5 -- 0.2    &2.0 -- 150.0    &   144   \cite{Adloff:2000qk}      &    1.7\%      & 1.0078        \\
H1 high  Q$^2$ 99--00 $e^+p$ NC    &0.002 -- 0.65    &100.0 -- 30000.0&   147   \cite{Adloff:2003uh}      &    1.5\%      & 0.9999        \\
H1       Q$^2$ 99--00 $e^+p$ NC    &5.0E-6 -- 0.02   &2.0 --  12.0    &   85   \cite{Collaboration:2009bp}&    0.5\%      & 0.9994        \\
H1       Q$^2$ 00     $e^+p$ NC    &2.0E-4 -- 0.1    &12.0 -- 150.0   &   99    \cite{Aaron:2009kv}       &    0.5\%      & 0.9987        \\
%--------------------------------------------------------------------------------
ZEUS SVX 95 $e^+p$ NC             &5.4E-5 -- 0.0019 &2.50 -- 17.0    &   30    \cite{Breitweg:1998dz}    &    1.5\%      & 1              \\
ZEUS 96--97 $e^+p$ NC             &6.32E-5 -- 0.65  &2.70 -- 30000.0 &   242   \cite{Chekanov:2001qu}    &    2\%        & 0.9995         \\
ZEUS 06--07 $e^+p$ NC             &5.0 E-4 -- 0.007 &20.0 -- 130.0   &   162   \cite{Chekanov:2009na}    &    2.7\%      & 1.0012         \\
ZEUS 98--99 $e^-p$ NC             &0.050 -- 0.65    &200.0 -- 30000.0&   92    \cite{Chekanov:2002ej}    &    1.8\%      & 0.9996         \\
ZEUS 99--00 $e^+p$ NC             &0.050 -- 0.65    &200.0 -- 30000.0&   90    \cite{Chekanov:2003yv}    &    2.5\%      & 0.9968         \\
H1   03--07 $e^{\pm}p$ NC         &2.9E-5 -- 0.01   &2.0  -- 120.0   &   134  \cite{Collaboration:2010ry}&    4  \%      & 0.9997         \\    \hline   \hline
%--------------------------------------------------------------------------------
%--------------------------------------------------------------------------------
NMC   $\mu p$ $F_2$               &0.0045 -- 0.5    &2.50 -- 65.0    &    126  \cite{Arneodo:1996qe}     &    2  \%      & 1.0023        \\
NMC   $\mu d$ $F_2$               &0.0045 -- 0.5    &2.50 -- 65.0    &    126  \cite{Arneodo:1996qe}     &    2  \%      & 1.0023        \\
NMC   $\mu n/\mu p$               &0.008 -- 0.675   &2.23 -- 99.03   &    156  \cite{Arneodo:1996kd}     &    0.15\%     & 1             \\
BCDMS $\mu p$ $F_2$               &0.07 -- 0.75     &7.5 -- 230.0    &    167  \cite{Benvenuti:1989rh}   &    3\%        & 0.9900        \\
BCDMS $\mu d$ $F_2$               &0.07 -- 0.75     &8.75 -- 230.0   &    155  \cite{Benvenuti:1989fm}   &    3\%        & 0.9900        \\
E665  $\mu p$ $F_2$               &0.0037 -- 0.38726&2.046 -- 64.269 &    53   \cite{Adams:1996gu}       &    1.85\%     & 1.0012        \\
E665  $\mu d$ $F_2$               &0.0037 -- 0.38726&2.046 -- 64.269 &    53   \cite{Adams:1996gu}       &    1.85\%     & 1.0012        \\
SLAC  $ep$ $F_2$                  &0.007 -- 0.65    &2.01 -- 22.21   &    53   \cite{Whitlow:1991uw}     &    2\%        & 1.0128        \\
HERMES $e p$ $F_2$                &0.006 -- 0.66    &2.0 -- 12.0     &    39   \cite{Airapetian:2011nu}  &    7.55\%     & 1.0167        \\
HERMES $e d$ $F_2$                &0.006 -- 0.66    &2.0 -- 12.0     &    39   \cite{Airapetian:2011nu}  &    7.55\%     & 1.0167        \\     \hline   \hline
%--------------------------------------------------------------------------------
%--------------------------------------------------------------------------------
H1    $ep$ $F_2^{c}$              &1.3E-4 -- 0.02   &6.50 -- 60.0    &    25   \cite{Aaron:2011gp}       &    7.5\%      & 1               \\
H1    $ep$ $F_2^{c}$              &0.005 -- 0.032   &200.0 -- 650.0  &    4    \cite{Aktas:2004az}       &    1.5\%      & 1               \\
H1    $ep$ $F_2^{c}$              &2.0E-4 -- 0.05   &5.0 -- 2000.0   &    29   \cite{:2009ut}            &    1.5\%      & 1               \\
H1    $ep$ $F_2^{c}$              &1.97E-4 -- 0.05  &12.0 -- 60.0    &    6    \cite{Aktas:2005iw}       &    1.5\%      & 1               \\
H1    $ep$ $F_2^{c}$              &1.3E-4 -- 0.00316&3.50 -- 60.0    &    10   \cite{Adloff:2001zj}      &    1.5\%      & 1               \\
H1    $ep$ $F_2^{c}$              &8.0E-4 -- 0.008  &12.0 -- 45.0    &    9    \cite{Adloff:1996xq}      &    1.5\%      & 1               \\
%--------------------------------------------------------------------------------
ZEUS  $ep$ $F_2^{c}$              &1.3E-4 -- 0.00676&4.20 -- 111.8   &    5    \cite{Chekanov:2007ch}    &    1.8\%      & 1               \\
ZEUS  $ep$ $F_2^{c}$              &1.3E-4-- 0.02    &4.0 -- 130.0    &    18   \cite{Breitweg:1999ad}    &    1.65\%     & 1               \\
ZEUS  $ep$ $F_2^{c}$              &3.0E-5 -- 0.03   &2.0 -- 500.0    &    31   \cite{Chekanov:2003rb}    &    2.2\%      & 1               \\
ZEUS  $ep$ $F_2^{c}$              &8.0E-5 -- 0.03   &30.0 -- 1000.0  &    8    \cite{Chekanov:2009kj}    &    1.5\%      & 1               \\
%--------------------------------------------------------------------------------
H1    $ep$ $F_2^{b}$              &0.005 -- 0.032   &200.0 -- 650.0  &    4    \cite{Aktas:2004az}       &    1.5\%      & 1               \\
H1    $ep$ $F_2^{b}$              &2.0E-4 -- 0.05   &5.0 -- 2000.0   &    12   \cite{:2009ut}            &    1.5\%      & 1               \\
H1    $ep$ $F_2^{b}$              &1.97E-4 -- 0.05  &12.0 -- 60.0    &    6    \cite{Aktas:2005iw}       &    1.5\%      & 1               \\
ZEUS  $ep$ $F_2^{b}$              &2.0E-4 -- 0.013   &12.0 -- 600.0  &    9    \cite{Abramowicz:2011kj}  &    2 \%       & 1               \\
ZEUS  $ep$ $F_2^{b}$              &1.3E-4 -- 0.013   &3.0 -- 450.0   &    11   \cite{Abramowicz:2010zq}  &    2 \%       & 1               \\
ZEUS  $ep$ $F_2^{b}$              &8.0E-5 -- 0.03   &30.0 -- 1000.0  &    8    \cite{Chekanov:2009kj}    &    1.5\%      & 1               \\
%--------------------------------------------------------------------------------
H1    $ep$ $F_2^{c}/F_2^{p}$      &8.0E-4 -- 0.008  &12.0 -- 45.0    &    9    \cite{Adloff:1996xq}      &    1.5\%      & 1               \\         \hline   \hline
%--------------------------------------------------------------------------------
%--------------------------------------------------------------------------------
FNAL E866/NuSea                   &0.026 -- 0.315   &54 (Fixed)      &    30  \cite{Towell:2001nh,Webb:2003ps} &    0.6\%      & 1          \\   \hline
%--------------------------------------------------------------------------------
H1/ZEUS $F_L$   &4.27E-5 -- 0.0049  &2.0 -- 110.0   &    127    \cite{Adloff:2000qk,Chekanov:2009na,Collaboration:2010ry,:2008tx}   &    -    & 1 \\   \hline
%--------------------------------------------------------------------------------
CHORUS $\nu N$ $xF_3$            &0.02 -- 0.65      &2.052 -- 81.55  &    50    \cite{Onengut:2005kv}   &    2.1\%            & 1.0023     \\
NuTeV $\nu N$ $xF_3$             &0.015 -- 0.75     &3.162 -- 125.89 &    64    \cite{Tzanov:2005kr}    &    2.1\%            & 1.0023     \\              \hline   \hline
%--------------------------------------------------------------------------------
D{\O} I $p\bar{p}$ incl.~jets          &   &    &    90    \cite{hep-ex/0011036}     &    6  \%            & 1.0021      \\
CDF II $p\bar{p}$ incl.~jets           &   &    &    76    \cite{Abulencia:2007ez}   &    5.8\%            & 1.0023      \\
D{\O} II $p\bar{p}$ incl.~jets         &   &    &    110   \cite{Abazov:2008hu}      &    6.1\%            & 1.0221      \\
ZEUS 96--97 $e^+p$ incl.~jets          &   &    &    30    \cite{Chekanov:2002be}    &    2  \%            & 0.9988      \\
ZEUS 98--00 $e^\pm p$ incl.~jets       &   &    &    30    \cite{Chekanov:2006xr}    &    2.5\%            & 0.9980      \\
CDF II $p\bar{p}$ incl.~jets           &   &    &    20    \cite{hep-ex/0512020}     &    6  \%            & 0.9972       \\       \hline
%--------------------------------------------------------------------------------
\textbf{All data sets }          &                 &                &         \textbf{ 3279 }     & &  \\                                                \hline   \hline
\end{tabular}}
}
\caption{Data sets fitted in our NLO QCD analysis. The fitted normalisations ${\cal N}_{n}$ of the data sets included in the global QCD fit, together with the total normalisation uncertainty, $\Delta {\cal N}_{n}$, for each data set $n$ are also shown in the table. The details of corrections to data and the kinematic cuts applied are contained in the text.}
\label{Table1}
\end{table*}
%
%=====================================================================================
%

Statistically, most significant data that we use in our QCD analysis,
which we call ``{\tt KKT12}" fit, are the HERA
measurements of the DIS `reduced' cross-section. We use the small and
large-$x$ from H1 and ZEUS $\sigma_{r}^\pm (x,Q^2)$ data for $Q^2\geq 2$ GeV$^2$
\cite{Adloff:1999ah,Adloff:2000qj,Adloff:2000qk,Adloff:2003uh,Collaboration:2009bp,Aaron:2009kv,
Breitweg:1998dz,Chekanov:2001qu,Chekanov:2009na,Chekanov:2002ej,Chekanov:2003yv}\,.

In our QCD fits, we use the most recent H1 and
ZEUS combined measurement for inclusive $e^{\pm}p$ scattering
cross sections \cite{:2009wt} instead of all the above mentioned data.
To distinguish, we call this ``{\tt KKT12C}" when we used the recent H1
and ZEUS combined measurement.

This combined data set contains complete information on DIS cross
sections published by the H1 and ZEUS collaborations. The
kinematic range of the neutral current data is 0.045\,GeV$^2$
$\leq Q^2 \leq$30000\,GeV$^2$ and 6$\times 10^{-7} \leq x \leq
$0.65, for values of inelasticity $y$ between $0.005$ and $0.95$.
The statistical uncertainty reduces in this measurement.

The most recent HERA structure function data which we use in our
analysis are the H1 measurements of the $e^{\pm} p$  DIS `reduced'
cross-section~\cite{Collaboration:2010ry}. The kinematic range of
the following measurement covers 1.5\,GeV$^2<Q^2<$120\,GeV$^2$ and
2.9$\times 10^{-5}<x<0.01$. By using these data, we can
improve our PDFs, especially at small $x$ value.

In addition, we have used the fixed target $F_2^p$ data of NMC~
\cite{Arneodo:1996qe}, BCDMS~\cite{Benvenuti:1989rh}, E665~
\cite{Adams:1996gu}, SLAC~\cite{Whitlow:1991uw} and HERMES~\cite{Airapetian:2011nu} all subject to
the standard cuts $Q^2 \geq 2$ GeV$^2$ and $W^2\! = Q^{2}
(\frac{1}{x} - 1) + m_{p} \! \geq\!12.5$ GeV$^2$. Furthermore the
fixed target data for $F_2^d$~\cite{Arneodo:1996qe,Adams:1996gu,Benvenuti:1989fm,Airapetian:2011nu} and the
structure function ratio $F_2^d/F_2^p$~\cite{Arneodo:1996kd} are both
subject to the same standard cuts have been added to the present
analysis. These data on $F_2^p$, $F_2^d$ and $F_2^d/F_2^p$ help us
to control the behavior of PDFs  in large $x$, which dominated by
valence quarks~\cite{DeRoeck:2011na}.

The H1 and ZEUS measurements on the heavy flavour contribution to
structure functions $F_2^c$, $F_2^b$ and $F_2^c/F_2^p$ of~
\cite{Aaron:2011gp,Aktas:2004az,:2009ut,Aktas:2005iw,Adloff:2001zj,Adloff:1996xq,Chekanov:2007ch,Breitweg:1999ad,Chekanov:2003rb,Abramowicz:2011kj,Abramowicz:2010zq,Chekanov:2009kj}
have also been included in the both sets of NLO fits. The
Drell--Yan dimuon pair production data of the E866/NuSea (fixed
target) experiment~\cite{Towell:2001nh,Webb:2003ps} have been
used.

The direct HERA measurements of the longitudinal structure function, $F_L$~
\cite{Adloff:2000qk,Chekanov:2009na,Collaboration:2010ry,:2008tx}
which is a direct constraint on the small $x$ region of gluon
distribution, have also been included in the present analysis. And
finally, we include the $\nu(\bar{\nu})N$ $xF_3$ data from CHORUS~
\cite{Onengut:2005kv} and NuTeV~\cite{Tzanov:2005kr} with the same
cuts in the present analysis. The importance of $xF_3$ data is that
the small difference between NC $e^{+}p$ and $e^{-}p$ cross
sections data at high Q$^2$, which determines $xF_3$ as seen in
Eq. \ref{Sigma}, contain the valence quark distributions. The
reason is that the structure function $xF_3$ is related to valence
quark contributions and it can provide direct tests of the total valence quark distribution.

We now indicate that we use the deuteron structure function
data in our QCD analysis also. Experimental data on the deuteron structure
function $F_2^d$ and the structure function ratio $F_2^d/F_2^p$
have to be corrected for nuclear shadowing effects~
\cite{Badelek:1994qg,Qiu:1986wh,Badelek:1991qa}. Since the
neutron DIS data come from the experiments with nuclear targets
such as deuteron, these effects have to be taken into account.
Taking the shadowing into account, the deuteron structure function
$F_2^d (x,Q^2)$ is related in the following way to $F_2^p(x,Q^2)$,
$F_2^n(x,Q^2)$ and to the shadowing contribution $\delta F_2^d (x,Q^2)$,

\begin{equation}
  F_2^d (x,Q^2) = \frac {F_2^p (x,Q^2) + F_2^n (x,Q^2)} {2} -  \delta F_2^d (x,Q^2) \,.
\end{equation}

We study the shadowing effect as a nuclear effect in our QCD fits and we found that in
the range of 0.1$<x<$0.75 for the NMC, BCDMS and E665  $F_2^d$ and
$F_2^d/F_2^p$ data, the shadowing effect are very small~\cite{Badelek:1991qa}.

The HERA~\cite{Chekanov:2002be,Chekanov:2006xr} and Tevatron~\cite{hep-ex/0011036,Abulencia:2007ez,Abazov:2008hu,hep-ex/0512020,arXiv:0807.2204}
Run I \& II data on inclusive jet production
in $e^+p$ and $p \bar p$ scattering have been used in our QCD analysis to extract the new sets of
parton distribution functions. Take into account the jet production data, constrains the
the mid- to high-$x$ ($x \approx 0.01 - 0.5$) gluon density and allows an accurate extraction of strong coupling constant $\alpha_s(M_Z^2)$ at NLO.

The {\tt fastNLO} package~\cite{Kluge:2006xs}, based on {\tt NLOJET++}~\cite{Nagy:2001fj,Nagy:2003tz}, presents
a method that offers an exact and very fast pQCD calculation for a large number of jet data sets. This scenario
also allows the inclusion of the NLO hard cross section corrections to both the Tevatron and HERA jet data in
the fitting procedure.  In the present PDF fits, we use the {\tt fastNLO} code to take full advantage of direct sensitivity of the
jet production data to the gluon density in the proton.

The total data sets that we use in the present global analysis are
listed in Table~\ref{Table1} and are ordered according to the type
of process. The fitted normalization ${\cal N}_{n}$ of the data
sets included in the global fit, together with the total
normalization uncertainty, $\Delta {\cal N}_{n}$, for each data
set $n$ are also shown in this table. There are 3279  and 2485 data
points for {\tt KKT12} and  {\tt KKT12C} fits, respectively.
We note that the data from Refs.~\cite{Collaboration:2010ry} to~\cite{Tzanov:2005kr}
are used in both of {\tt KKT12} and  {\tt KKT12C} QCD fits.
It can be a new feature of our analysis that we use a wide range of the
most recent DIS data sets including different processes, NC,
$e^+p$ and $e^-p$ scattering to extract all PDFs.

%
%%%%%%%%%%%%%%%%%%%%%%%%%%%%%%%%%%%%%%%%%%%%%%%%%%%%%%%%%%%%%%%%%%%%%%%%%%%%%%%%%%%%%%%%%%%%%%%%%%%%%%%%%
%

\subsection{The method of the minimization and error calculation}\label{The method of the minimization}

%
%-----------------------------------------------------------
%

\subsubsection{ Minimization of $\chi^2$  }

The global goodness-of-fit procedure follows the usual chi--squared method with $\chi^2_{\rm global} (p)$ defined as~
\cite{Martin:2009iq,Khorramian:2009xz,Khorramian:2008yh,Stump:2001gu}
\begin{eqnarray} \label{chi2}
\chi_{\mathrm{\rm global}}^{2}(p) &=& \sum_{i=1}^{n^{\rm data}} \left[\left(\frac{{\cal N}_{i} - 1}{\Delta{\cal N}_{i}}\right)^{2} \right. \nonumber \\ \left.
\right. &+& \left. \sum_{j=1}^{n^{\rm data}} \left( \frac{{\cal N}_{j} D_{j}^{\rm data} - T^{\rm theory}_{j} (p)} {{\cal N}_{j} ~ \Delta D_{j}^{\rm data}} \right)^{2}\right]\,,
\end{eqnarray}
where $p$ denotes the set of $13$ independent parameters in the
fit, including $\alpha_s(Q_0^2)$, and $n^{data}$ is the number of
data points included, $n^{data} \,=\, 3279$ ({\tt KKT12}) and $2485$
({\tt KKT12C}) for the NLO fit. The errors include systematic and
statistical uncertainties, being the total experimental error
evaluated in quadrature. For the $i^{\mathrm{th}}$ experiment,
$D_{i}^{\rm data}$, $\Delta D_{i}^{\rm data}$, and $T_{,i}^{\rm theory}$
denote the data value, measurement uncertainty (statistical and
systematic combined), and theoretical value for the
$n^{\mathrm{th}}$ data point for proton structure function as an example. The theory prediction $T_{i}^{\rm theory}$ depends on the input PDF parameters $p$.
The {\tt KKT12} and {\tt KKT12C} fits added all uncorrelated and correlated experimental errors in quadrature in the $\chi^2_{\rm global}$ definition.
${\Delta{\cal N}_{n}}$ is the experimental normalization uncertainty and ${\cal N}_{n}$ is an
overall normalization factor for the data of experiment $n$.

In the present analysis, we allow the normalization of several data sets to be free at the same
time as the PDF parameters, and then fixed them in the determination of the PDF uncertainties.
We also allow a different normalization for each of the HERA running periods.
To determine the best fit values at NLO, we minimize the $\chi_{\rm global}^2$ with respect to 27 free input PDF parameters,
together with $\alpha_S(Q_0^2)$ and 23 different data set normalizations, giving a total of 50 free parameters in the fits.
The minimization of the above $\chi ^2$ value to determine the best
parametrization of the parton distributions is done using the
program {\tt MINUIT} \cite{MINUIT}.
Due to the limitation of {\tt MINUIT} package in the practical maximum number of free parameters, we fixed
some of the data set normalization to be 1. For example, in the {\tt KKT12}  fit, we take all the normalization
of the H1 and ZEUS data on $F_2^{\bar {c} c} (x,Q^2)$ to be fixed at 1. (See Table~\ref{Table1}.)
As discussed in Sec.~\ref{data-sets}, we use the H1/ZEUS combined data in our {\tt KKTC} fit,
consequently the number of free normalization decreases
so we have this chance to allow the $F_2^{\bar {c} c} (x,Q^2)$ data to have
different normalization.

In the global fits, we obtain
\begin{eqnarray} \label{chi2valueNLO}
\frac {\chi^{2}}{\rm NDF} = \frac{3590.589}{3266} &=& 1.098 \, {\tt (KKT12) } \, ,  \nonumber \\
\frac{2622.118}{2472} &=& 1.060  \, {\tt (KKT12C) }  \, ,
\end{eqnarray}
with the total data sets in Table~\ref{Table1} and for the parameter values listed in Table~\ref{Table2}.
%Above obtained $\chi^{2}$ show that the Combined HERA data have a slightly better fit quality at NLO.
Consequently the $\chi ^2$ fit value for {\tt KKT12} is almost equal to {\tt KKT12C} QCD fit at NLO.

%
%-----------------------------------------------------------
%
\subsubsection{ Error calculation using Gaussian method}

The one $\sigma$ errors for the parton density $x f_i (x,Q^2)$ are given by
the Gaussian error propagation method as discussed in details in Refs.~\cite{Bluemlein:2002be,Pumplin:2001ct,Khorramian:2010qa}.
In this method, the standard linear errors propagation allows one to calculate the error
on any quantity $F (p_i)$ using the formula
\begin{equation}
\Delta F = \biggl[\, \Delta\chi^2 \sum_{i=1}^{n}\sum_{j=1}^n \frac{\partial F}{\partial p_i} C_{ij}(p)\frac{\partial F}{\partial p_i} \, \biggr]^{1/2}\, ,
\end{equation}
where $\Delta \chi^2$ is the allowed variation in $\chi^2$ and $C_{ij}(p) =  (H^{-1})_{ij}$ is the covariance, or error
matrix of the parameters determined in the QCD analysis at the scale $Q_0^2$  and  $n=13$ is
the number of free input parameters in the global fit (see Table~\ref{covmat}).

\begin{sidewaystable}
\centering
%\vspace{160mm}
% {\footnotesize
{\small
\begin{tabular}{||c||c|c|c|c|c|c|c|c|c|c|c|c|c|c|c|c||}
\hline \hline
%        &  &  &  &  &  &  &  &  &  &&&&& \\
& $\alpha_u$ & $\beta_u$ & $\alpha_d$ & $\beta_d$ & $A_{\Delta}$ & $\alpha_{\Delta}$ &  $\beta_{\Delta}$ & $A_S$ & $\alpha_S$ & $\beta_S$ & $\alpha_g$ & $\beta_g$ &$\alpha_{s}(Q_0^2)$  \\

\hline \hline
 $\alpha_u$   & 2.39$\times 10^{-5}$ &  &  &  &  &  &  & &&&&& \\
\hline
 $\beta_u$    & 7.20$\times 10^{-5}$  &  2.36$\times 10^{-4}$&  &  &  &  &  & &&&&& \\
\hline
 $\alpha_d$   & -6.30$\times 10^{-5}$ & -1.86$\times 10^{-4}$ & 2.22$\times 10^{-4}$ &  &  &  &  &&&&&&  \\
\hline
 $\beta_d$    &-1.68$\times 10^{-4}$  & -5.30$\times 10^{-4}$ &   7.11$\times 10^{-4}$&   3.01$\times 10^{-3}$ &  &  &  & &&&&& \\
\hline
 $A_{\Delta}$  &-6.86$\times 10^{-7}$ &  -2.19$\times 10^{-6}$  &  2.65$\times 10^{-6}$  &  8.74$\times 10^{-6}$  & 6.54$\times 10^{-4}$   &  &  & &&&&& \\
\hline
 $\alpha_{\Delta}$ & -4.51$\times 10^{-5}$ & -1.40$\times 10^{-4}$ & 1.38$\times 10^{-4}$ & 4.33$\times 10^{-4}$ & -1.21$\times 10^{-5}$ & 2.38$\times 10^{-4}$ &  & &&&&& \\
\hline
 $\beta_{\Delta}$    & -1.01$\times 10^{-3}$ & -2.96$\times 10^{-3}$ & 2.95$\times 10^{-3}$ & 8.07$\times 10^{-3}$ & -1.10$\times 10^{-4}$ & 4.56$\times 10^{-4}$ & 7.91$\times 10^{-2}$ & &&&&& \\
\hline
 $A_S$    & 3.62$\times 10^{-6}$ &  1.06$\times 10^{-5}$ &  -1.56$\times 10^{-5}$ & -5.40$\times 10^{-5}$ & -1.72$\times 10^{-7}$ & -9.29$\times 10^{-6}$ & -1.92$\times 10^{-4}$ &  1.41$\times 10^{-6}$ &&&&& \\
\hline
 $\alpha_S$    & 2.88$\times 10^{-6}$ & 8.76$\times 10^{-6}$ &  -7.16$\times 10^{-6}$ & -1.89$\times 10^{-5}$ & 2.18$\times 10^{-7}$ & -6.40$\times 10^{-6}$ &  -1.26$\times 10^{-4}$ & 5.11$\times 10^{-7}$ & 1.17$\times 10^{-6}$ &&&& \\
\hline
 $\beta_S$ & 9.96$\times 10^{-5}$ & 3.16$\times 10^{-4}$ & -6.12$\times 10^{-5}$ &  1.60$\times 10^{-4}$ & 3.98$\times 10^{-6}$ &  -1.72$\times 10^{-4}$ & -2.89$\times 10^{-3}$ & -9.68$\times 10^{-6}$ & 2.53$\times 10^{-5}$ & 1.85$\times 10^{-3}$ &&& \\
\hline
 $\alpha_g$    & 6.01$\times 10^{-5}$ &  1.81$\times 10^{-4}$ &  -1.66$\times 10^{-4}$ &  -4.62$\times 10^{-4}$ & -2.07$\times 10^{-6}$ & -1.22$\times 10^{-4}$ &  -2.52$\times 10^{-3}$ & 9.89$\times 10^{-6}$ & 6.74$\times 10^{-6}$ & 2.47$\times 10^{-4}$ & 1.67$\times 10^{-4}$ && \\
\hline
 $\beta_g$    & 4.23$\times 10^{-4}$ & 1.33$\times 10^{-3}$ &  -1.09$\times 10^{-3}$ & -2.72$\times 10^{-3}$ &  -9.46$\times 10^{-6}$ & -8.13$\times 10^{-4}$ &  -1.77$\times 10^{-2}$ &  6.13$\times 10^{-5}$ & 5.88$\times 10^{-5}$ & 2.18$\times 10^{-3}$ & 1.16$\times 10^{-3}$ & 1.07$\times 10^{-2}$ & \\
\hline
 $\alpha_{s}(Q_0^2)$ & 7.54$\times 10^{-6}$ &  2.23$\times 10^{-5}$ & -1.86$\times 10^{-5}$ & -5.08$\times 10^{-5}$ & -1.50$\times 10^{-7}$ &  -1.47$\times 10^{-5}$ &  -3.01$\times 10^{-4}$ &  1.04$\times 10^{-6}$ & 1.02$\times 10^{-6}$ & 3.54$\times 10^{-5}$ & 2.04$\times 10^{-5}$ & 1.39$\times 10^{-4}$ & 2.91$\times 10^{-6}$ \\
\hline \hline
\end{tabular}
}
\caption{The covariance matrix for the $12 + 1$ parameters of {\tt KKT12} NLO fit, based on recent DIS world data.}\label{covmat}
\end{sidewaystable}

Consequently one can calculate the error of any quantity by using Hessian or covariance matrix.
This method recently used in our nucleon polarized and un-polarized structure function QCD analysis~\cite{Khorramian:2010qa,LightCone2011,EPS-HEP2011}.
The covariance matrix elements for 13 parameters in the {\tt KKT12} fit are given in Table~\ref{covmat}.
We able to calculate the uncertainties of PDFs using these covariance matrix
elements based on the Gaussian method as mentioned in this section. Their value at $Q^2$ are calculated by the QCD evolution.

%
%%%%%%%%%%%%%%%%%%%%%%%%%%%%%%%%%%%%%%%%%%    Discussion of fit results    %%%%%%%%%%%%%%%%%%%%%%%%%%%%%%%%%%%%%%%%%%%%%%%%%
%
\section{Discussion of fit results} \label{fit-results}
Both {\tt KKT12} and {\tt KKT12C} fits show acceptable agreements
with the data sets listed in Table~\ref{Table1}. In
order to discuss the fit results, we will concentrate on {\tt
KKT12} fit. Since the outcome of the {\tt KKT12} and {\tt KKT12C} for
$\chi ^2$ is almost similar and for PDFs is slightly different, we will show the
results of both QCD fits in some figures separately.

%
%-----------------------------------------------------------
%
\subsection{ Detailed comparison to the HERA and fixed target data}

\begin{figure*}
\vspace{1cm}
\centerline{\includegraphics[width=0.75\textwidth]{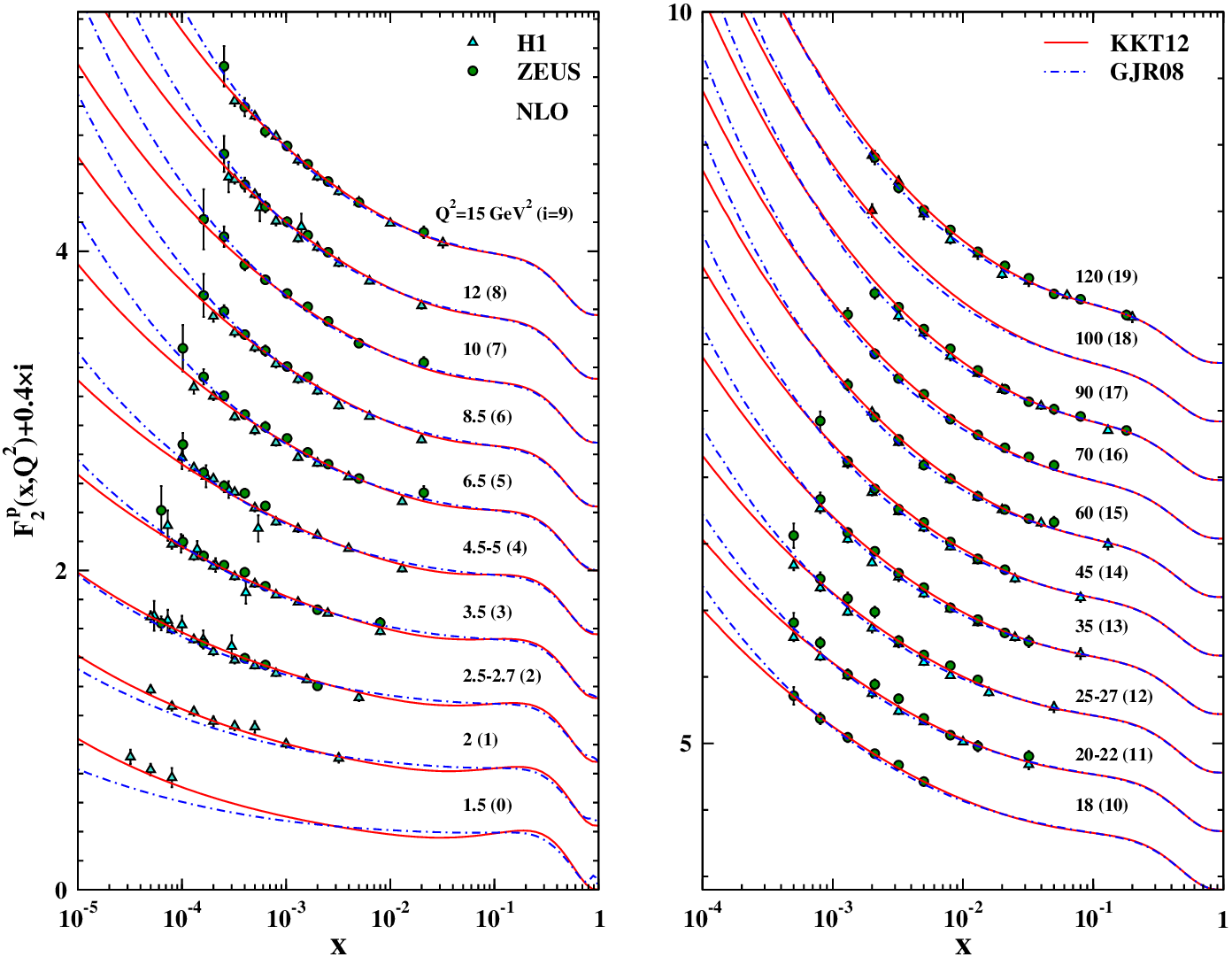}}
\caption{Comparison of our standard NLO\,($\overline{\rm
MS}$) results for $F_2^{p}(x,Q^2)$ with HERA data
\cite{Adloff:1999ah,Adloff:2000qj,Adloff:2000qk,Adloff:2003uh,Breitweg:1998dz,Chekanov:2001qu}
and Dortmund group, GJR08, \cite{Gluck:2007ck}. To facilitate the
graphical presentation we have plotted $F_2^p(x,Q^2)+0.4\times
i(Q^2)$ with $i(Q^2)$ indicated in parentheses in the figure.}
\label{QCDFit1}
\vspace{1.0cm}
\centerline{\includegraphics[width=0.75\textwidth]{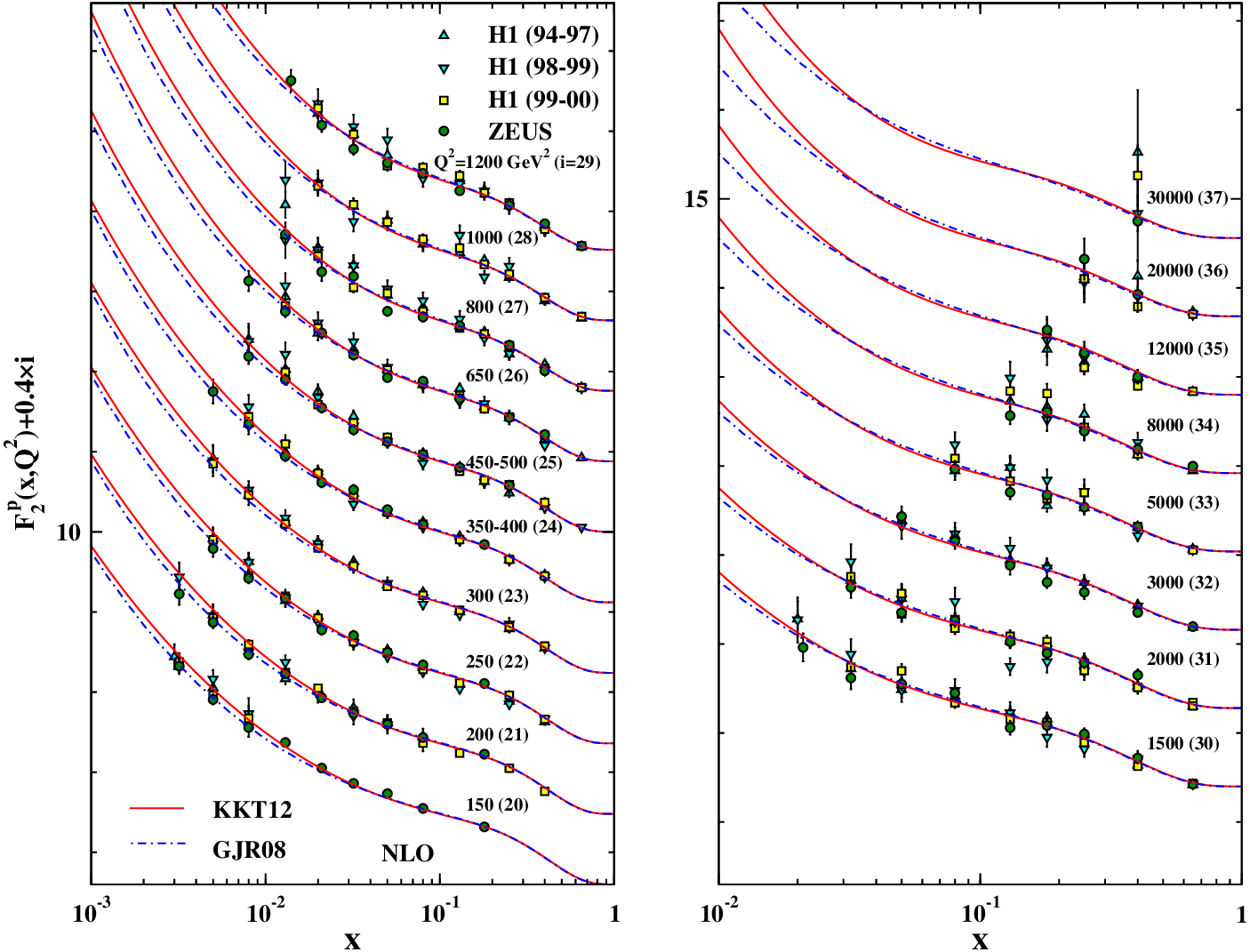}}
\caption{ As in previous figure but for large values of Q$^2$ and larger $x$.}
\label{QCDFit2}
\end{figure*}

Representative comparisons of our results with the relevant HERA
(H1 and ZEUS) data \cite{Adloff:1999ah,Adloff:2000qj,Adloff:2000qk,Adloff:2003uh,Breitweg:1998dz,Chekanov:2001qu}
on the structure function of the proton, $F_2^p(x,Q^2)$, are
presented in Figs.~\ref{QCDFit1} and \ref{QCDFit2}. As shown, the
data are well described throughout the whole medium-- to
small--$x$ region for $Q^2\!\gtrsim\!$ 1.5 GeV$^2$, and thus
perturbative QCD is here fully operative. It should be emphasized
that the perturbatively stable QCD predictions are in agreement
with all recent high statistics measurements of the $Q^2$-dependance of $F_{2} (x,Q^2)$ in wide range of $x$.
%Overall, a very good agreement is obtained.

%
%=====================================================================================
%
%
%=====================================================================================
%

\begin{figure*}
\vspace{1cm}
\centerline{\includegraphics[width=0.83\textwidth]{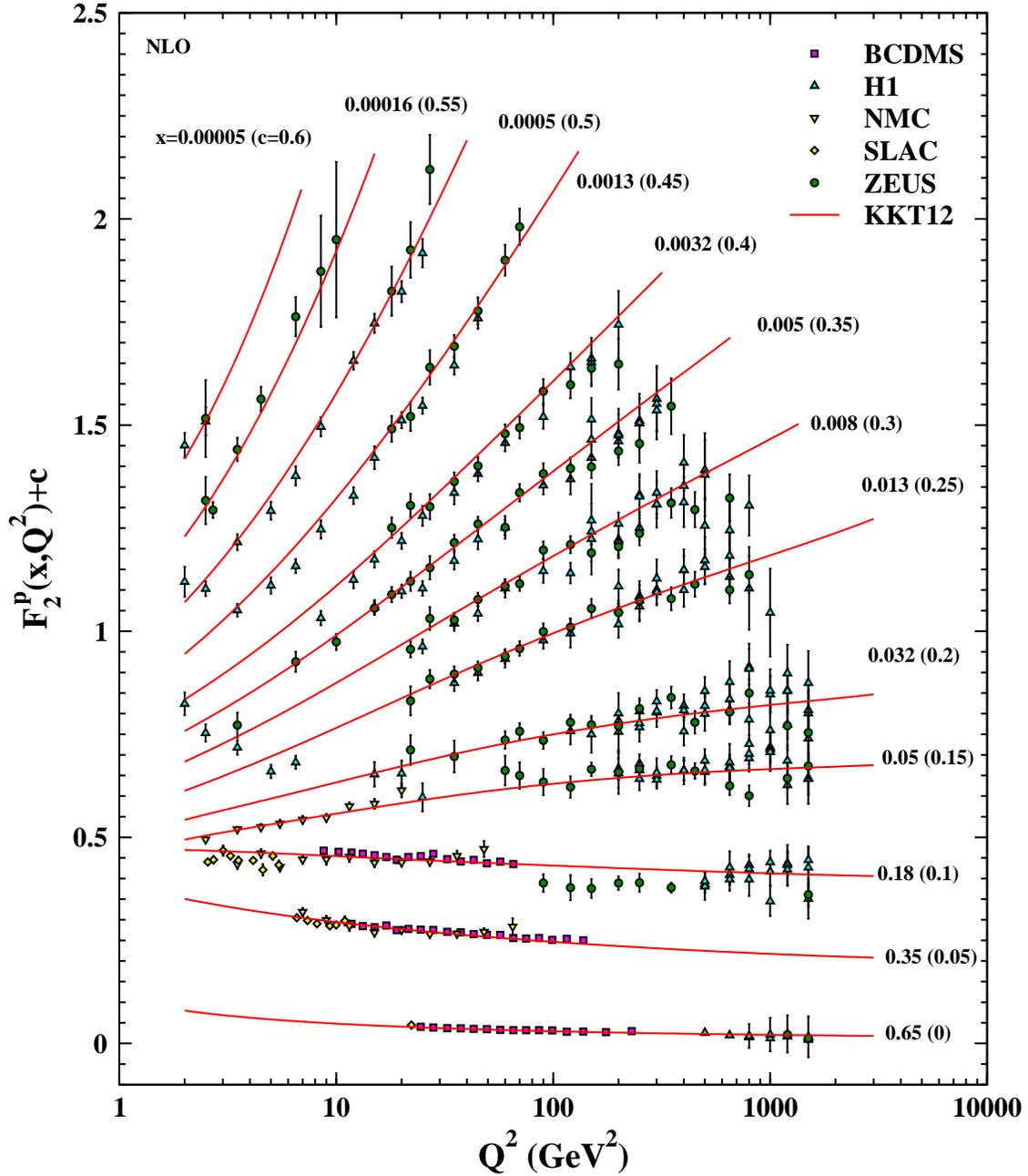}}
\caption{Comparison of our standard NLO\,($\overline{\rm
MS}$) results for $F_2^{p}(x,Q^2)$ with HERA data
\cite{Adloff:1999ah,Adloff:2000qj,Adloff:2000qk,Adloff:2003uh,Breitweg:1998dz,Chekanov:2001qu}
and fixed target data of NMC \cite{Arneodo:1996qe}, BCDMS
\cite{Benvenuti:1989rh} and SLAC \cite{Whitlow:1991uw}. To facilitate
the graphical presentation we have plotted $F_2^p(x,Q^2)+c $ with
$c$ indicated in parentheses in the figure.}
\label{F2P}
\end{figure*}

A detailed comparison of our next-to-leading
order (NLO)\,($\overline{\rm MS}$) results for $F_2^{p}(x,Q^2)$
with a selection of HERA data
\cite{Adloff:1999ah,Adloff:2000qj,Adloff:2000qk,Adloff:2003uh,Breitweg:1998dz,Chekanov:2001qu}
and fixed target data of NMC \cite{Arneodo:1996qe}, BCDMS
\cite{Benvenuti:1989rh} and SLAC \cite{Whitlow:1991uw} are shown
in Fig.~\ref{F2P} To facilitate the graphical presentation we have plotted
$F_2^p(x,Q^2)+ c $ with $c$ indicated in parentheses in the figure.

%
%=====================================================================================
%
%
%=====================================================================================
%
%

\begin{figure*}
\vspace{1cm}
\centerline{\includegraphics[width=0.67\textwidth]{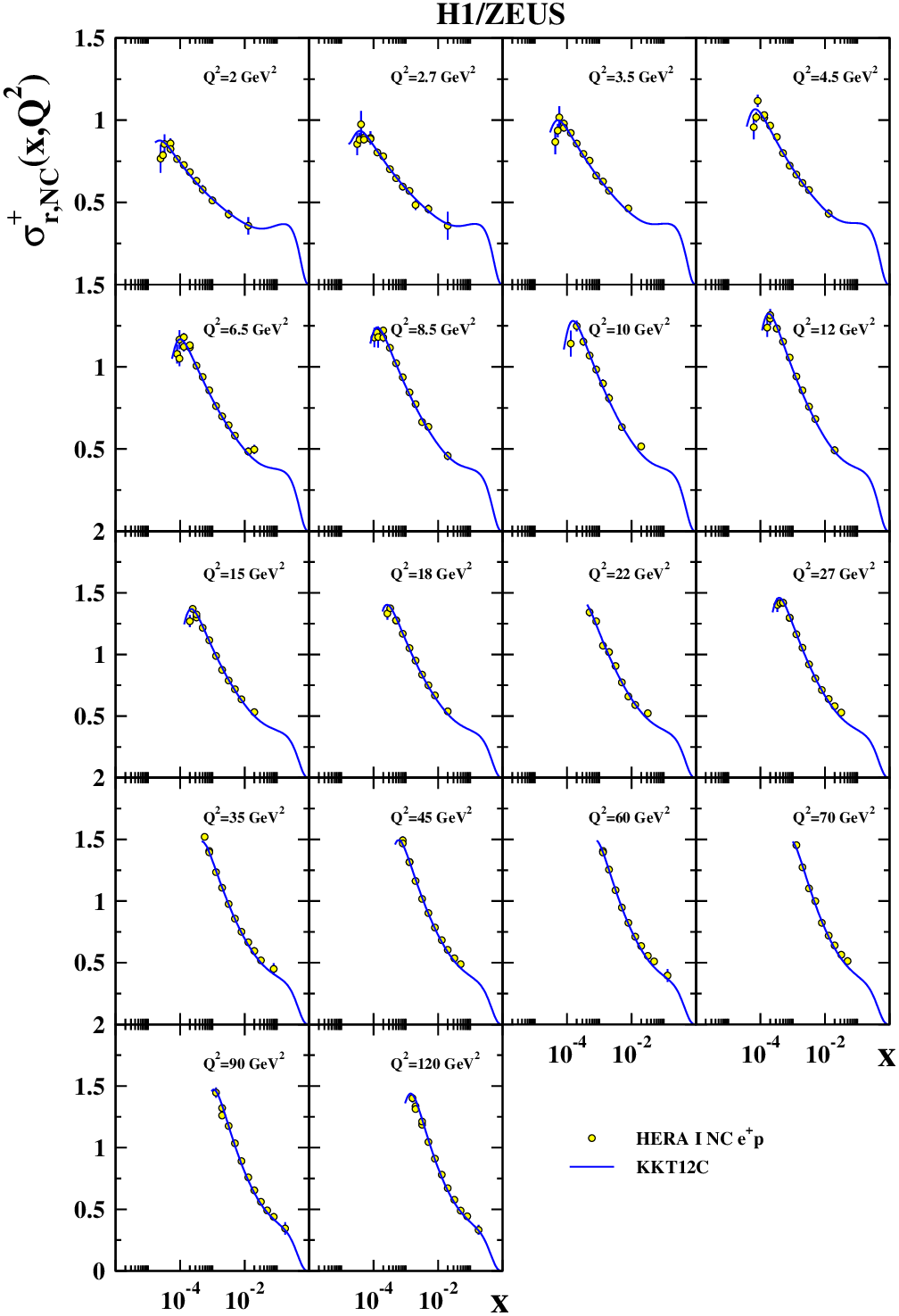}}
\caption{Our results of {\tt KKT12C} fit for reduced cross section, $\sigma^{+}_{r,NC} (x,Q^2)$, in comparison with HERA
combined NC $e^+ p$ reduced cross section \cite{:2009wt} as a function of $x$ for different
values of $Q^2$ bins for 2 GeV$^2$ $\leq$ Q$^2$ $\leq$ 120 GeV$^2$. The error bars indicate the total experimental uncertainty.}
\label{SigmaP-LowQ2}
\end{figure*}

\begin{figure*}
\vspace{1cm}
\centerline{\includegraphics[width=0.67\textwidth]{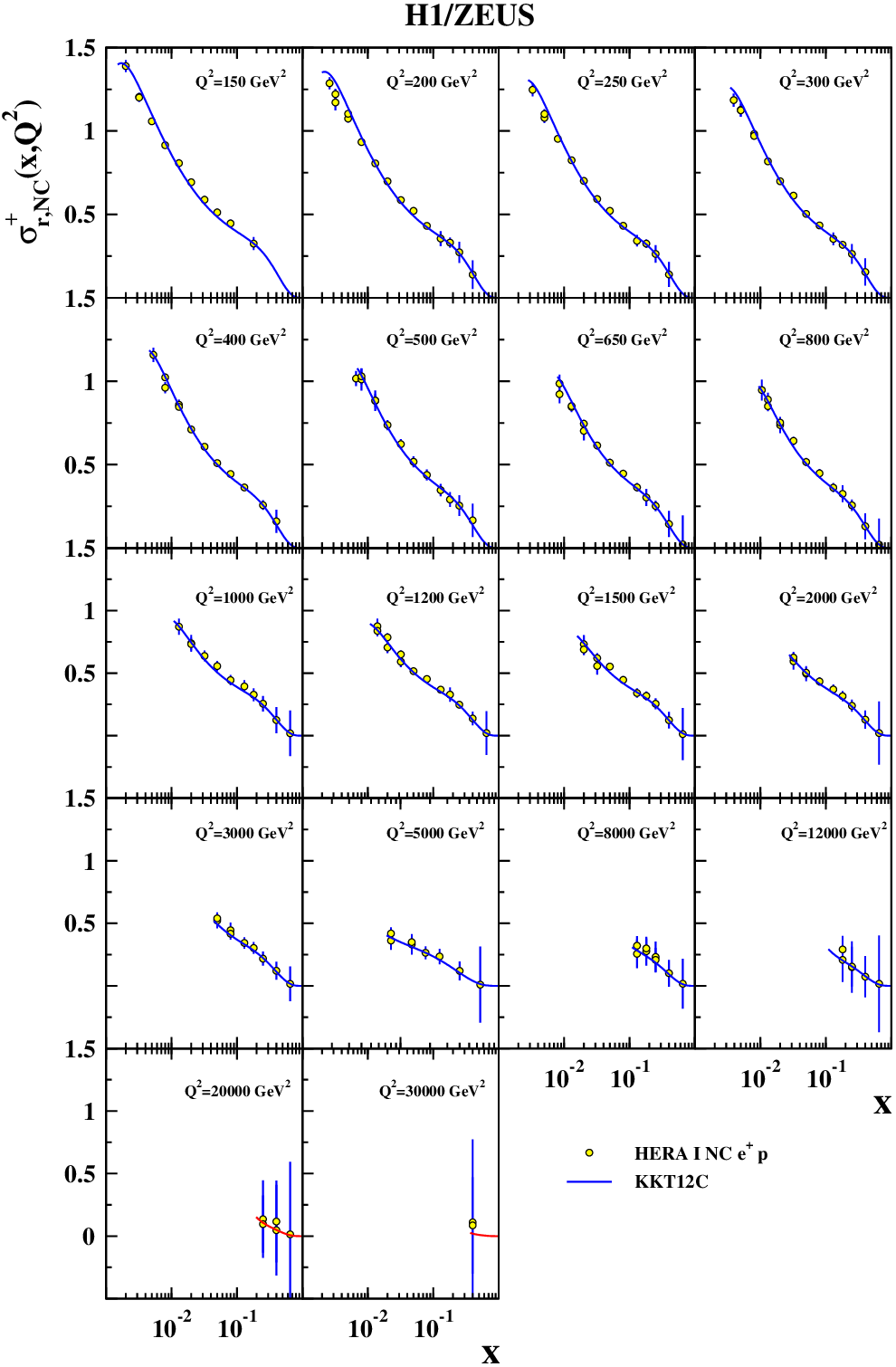}}
\caption{Our results of {\tt KKT12C} fit for reduced cross section, $\sigma^{+}_{r,NC} (x,Q^2)$, in comparison with HERA
combined NC $e^+ p$ reduced cross section \cite{:2009wt} as a function of $x$ for different
values of $Q^2$ bins for 150 GeV$^2$ $\leq$ Q$^2$ $\leq$ 30000 GeV$^2$. The error bars indicate the total experimental uncertainty.}
\label{SigmaP-HighQ2}
\end{figure*}

\begin{figure*}
\vspace{1cm}
\centerline{\includegraphics[width=0.67\textwidth]{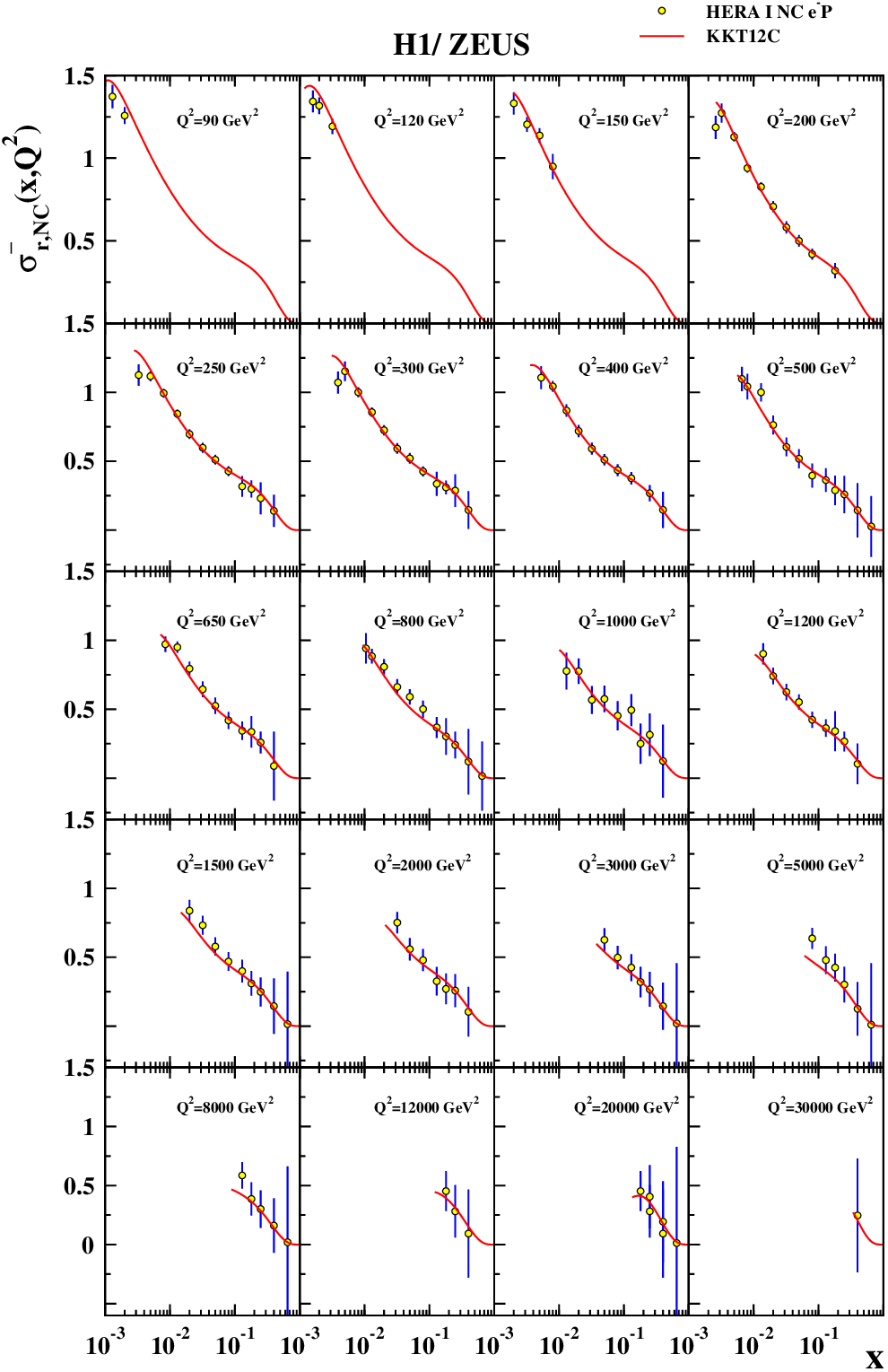}}
\caption{Our results of {\tt KKT12C} fit for reduced cross section, $\sigma^{-}_{r,NC} (x,Q^2)$, in comparison with HERA
combined NC $e^- p$ reduced cross section \cite{:2009wt} as a function of $x$ for different
values of $Q^2$ bins. The error bars indicate the total experimental uncertainty.}
\label{SigmaM}
\end{figure*}

A comparison of our {\tt KKT12C} theory predictions for the reduced cross section with the
H1/ZEUS combined NC $e^+ p$ and $e^- p$ data \cite{:2009wt} are shown in Figs.~\ref{SigmaP-LowQ2}, \ref{SigmaP-HighQ2} and \ref{SigmaM}.
The error bars in the figures indicate the total experimental uncertainty.
From the figures it is clear that our {\tt KKT12C} fits are in acceptable
agreement with the combined H1/ZEUS set of reduced DIS cross sections.

%
%
%=====================================================================================
%
%
%=====================================================================================
%
%
%
%=====================================================================================
%
%
%=====================================================================================
%
%
%=====================================================================================
%
%
%
%=====================================================================================
%
%

\begin{figure*}
\vspace{1cm}
\centerline{\includegraphics[width=0.85\textwidth]{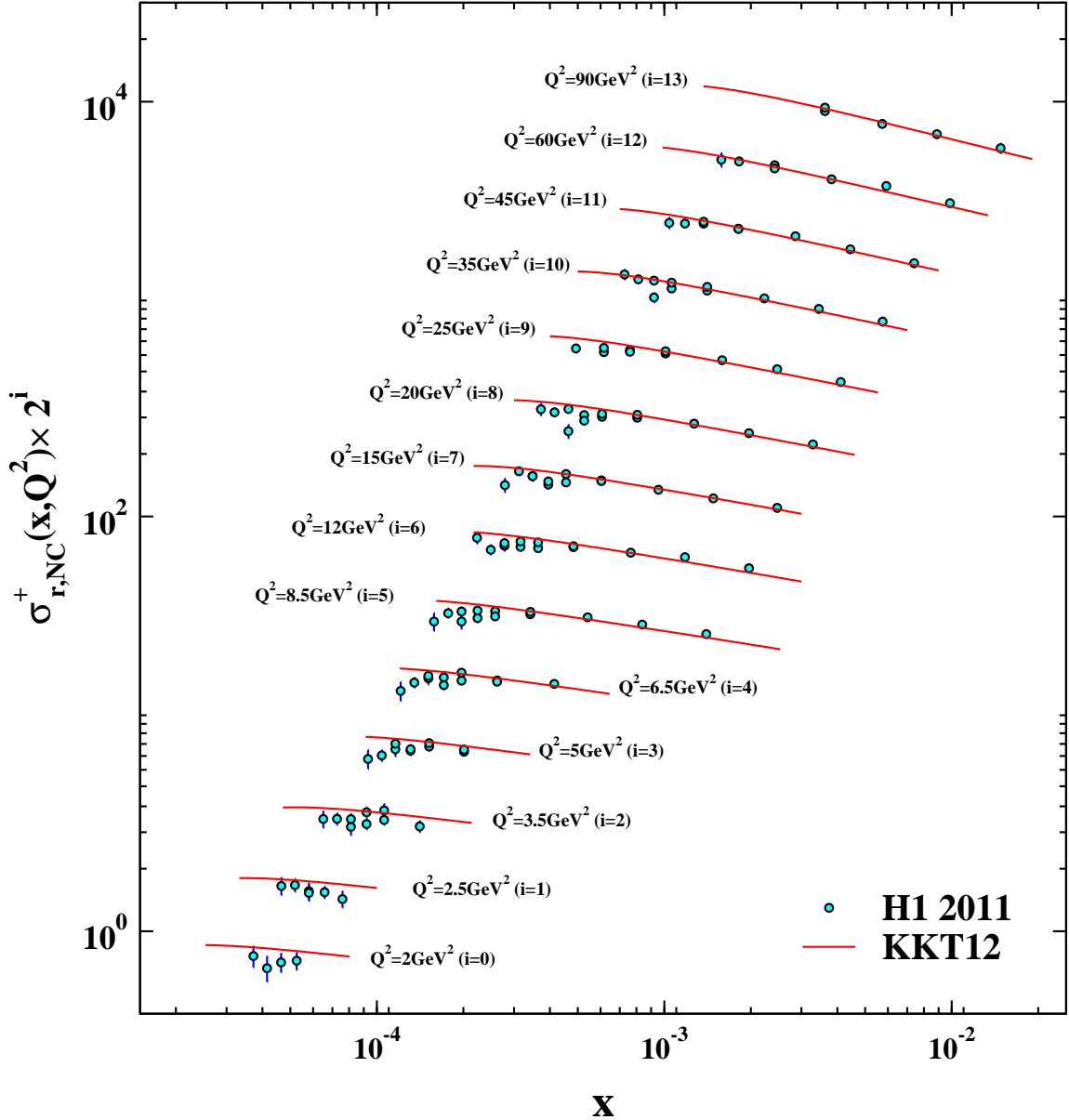}}
\caption{ Our results for reduced cross section, $\sigma_{r,NC} (x,Q^2)$, in comparison with the recent H1
NC $e p$ reduced cross section data \cite{Collaboration:2010ry}  as a function of $x$ for different
values of $Q^2$. }
\label{SigmaH12011}
\end{figure*}

Fig.~\ref{SigmaH12011} shows our prediction for reduced cross section, $\sigma_{r,NC} (x,Q^2)$
as a function of $x$ and for different value of $Q^2$, in comparison with the most recent data from H1 collaboration \cite{Collaboration:2010ry}.

%
%=====================================================================================
%
%
%=====================================================================================
%

%\clearpage

%
%-----------------------------------------------------------
%
\subsection{ KKT12 and KKT12C PDF sets}

The resulting parameters of the {\tt KKT12} and {\tt KKT12C} fits are summarized in Table~\ref{Table2}.
The $\chi^2/{\rm dof}$ for both PDFs fits also shown as well. As mentioned in Sec.~\ref{Parametrization}, the
values without errors have been fixed after a first minimization
since the data do not constrain these parameters well enough.
The parameter errors quoted are due to the propagation of the statistical and systematic errors in the data.

%
%=====================================================================================
%
\begin{table}[tbh]
%\vspace{1cm}
%\renewcommand{\arraystretch}{0.50}
\centering
{\begin{tabular}{|c|c|c|c|} \hline  \hline Parameter
&     Fit A - {\tt (KKT12) }   &      Fit B - {\tt (KKT12C) }                     \\
\hline  \hline
%--------------------------------------------------------------------------------
%--------------------------------------------------------------------------------
$A_u$                 & 0.3346                    & 0.3753                        \\
$\alpha_u$            & 0.3271  $\pm$ 0.0039      & 0.3126  $\pm$  0.0041         \\
$\beta_u $            & 3.5858  $\pm$ 0.0123      & 3.6536  $\pm$  0.0103         \\
$\delta_u$            & 0.4573                    & 0.5036                        \\
$\gamma_u$            & 5.4618                    & 3.2716                        \\
$\eta_u  $            & 20.1864                   & 21.2864
\\    \hline
%--------------------------------------------------------------------------------
%--------------------------------------------------------------------------------
$A_d$                 & 0.3730                    & 0.6124                        \\
$\alpha_d$            & 0.3668  $\pm$ 0.0105      & 0.4226  $\pm$  0.0092         \\
$\beta_d $            & 4.8277  $\pm$ 0.1001      & 5.1529  $\pm$  0.0654         \\
$\delta_d$            & 0.6306                    & 0.9553                        \\
$\gamma_d$            & 4.6633                    & -3.7475                       \\
$\eta_d  $            & 7.5520                    & 14.1380
\\    \hline
%--------------------------------------------------------------------------------
%--------------------------------------------------------------------------------
$A_ \Delta$           & 10.0048 $\pm$ 0.8417      & 15.8096 $\pm$  0.3160         \\
$\alpha_ \Delta$      & 1.7030  $\pm$ 0.0290      & 1.4430  $\pm$  0.0336         \\
$\beta_ \Delta $      & 9.8938  $\pm$ 0.3506      & 10.4343 $\pm$  0.4278         \\
$\delta_ \Delta$      & 0.6625                    & 0.4972                        \\
$\gamma_ \Delta$      &-7.3512                    & -7.5151                       \\
$\eta_ \Delta  $      & 23.7067                   & 20.4901
\\    \hline
%--------------------------------------------------------------------------------
$A_ S$                & 0.3494  $\pm$ 0.0019      & 0.3506  $\pm$   0.0023        \\
$\alpha_S$            & -0.1740 $\pm$ 0.0012      & -0.1775 $\pm$   0.0012        \\
$\beta_S $            & 8.2970  $\pm$ 0.0953      & 8.9898  $\pm$   0.0990        \\
$\delta_S$            & 0.5129                    & 0.8764                        \\
$\gamma_S$            & 0.2434                    & 0.4549                        \\
$\eta_S  $            & 9.8585                    & 12.4246
\\    \hline
%--------------------------------------------------------------------------------
%--------------------------------------------------------------------------------
$A_ g$               & 6.2512                    & 7.1383                        \\
$\alpha_g$           & 0.1179   $\pm$ 0.0101     & 0.1348   $\pm$ 0.0051         \\
$\beta_g $           & 8.4027   $\pm$ 0.1663     & 10.0188  $\pm$ 0.1494         \\
$\gamma_g$           & 0.2600                    & 1.3264                        \\
$\eta_g  $           &-1.4998                    & -0.9499
\\    \hline
%--------------------------------------------------------------------------------
%--------------------------------------------------------------------------------
$\chi^2/{\rm dof}$   &3590.589/3266 = 1.098      &  2622.118/2472 = 1.060
\\    \hline   \hline
%--------------------------------------------------------------------------------
\end{tabular}}
\caption{  Minimum values of $\chi^2$ together with the input PDFs parameters at
$Q_0^2 = 2$ GeV$^2$ determined from the two different global
analysis (Fit A  for {\tt KKT12} and Fit B for {\tt KKT12C} case).}
\label{Table2}
\end{table}
%
%=====================================================================================
%
%
%

\begin{figure}
\centerline{\includegraphics[width=0.65\textwidth]{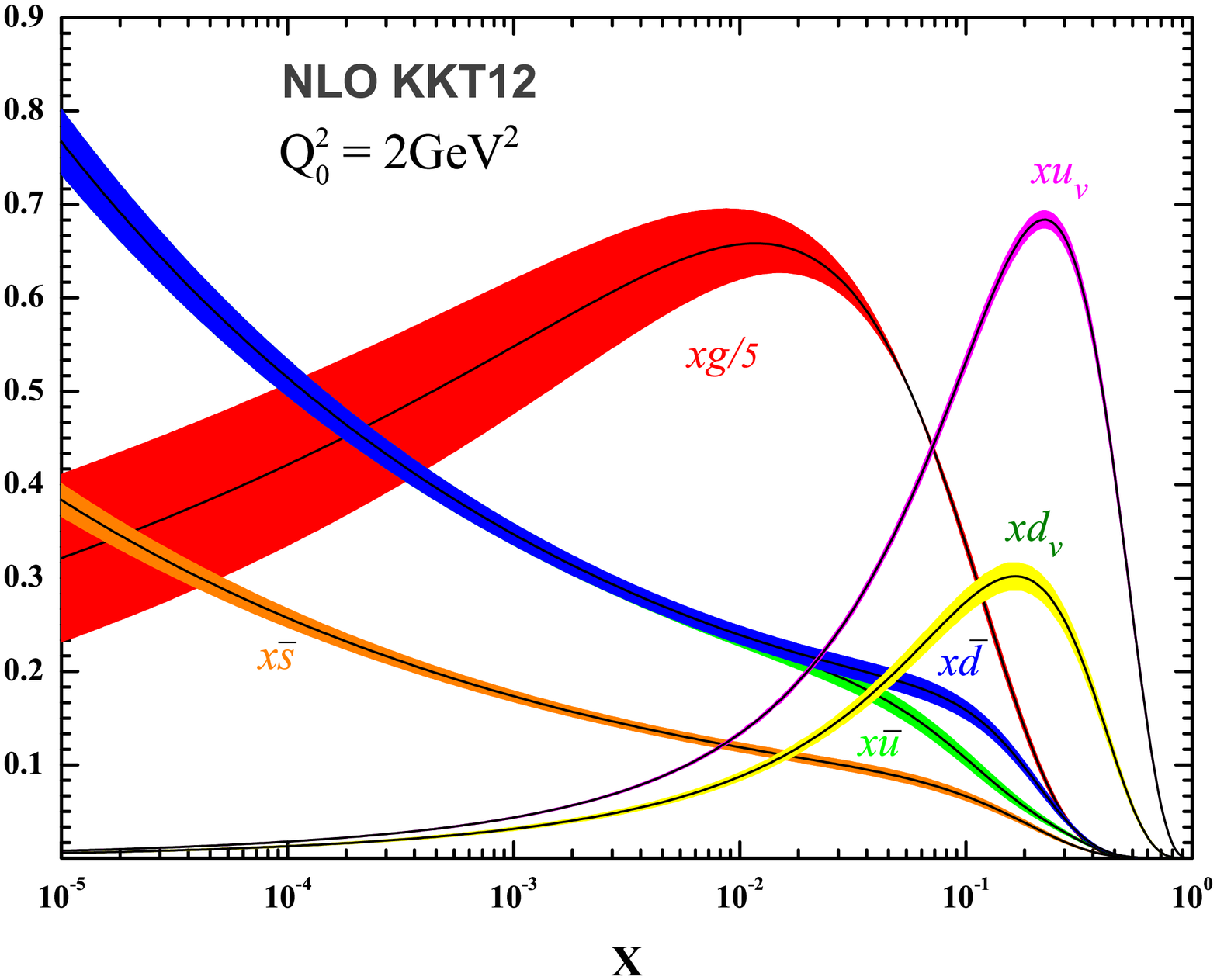}}
\caption{The {\tt KKT12} parton distributions at input scale
$Q_0^2$ = 2 GeV$^2$ as a function of $x$ in the NLO
approximation.} \label{NLOKKT12}
\end{figure}

\begin{figure}
\centerline{\includegraphics[width=0.65\textwidth]{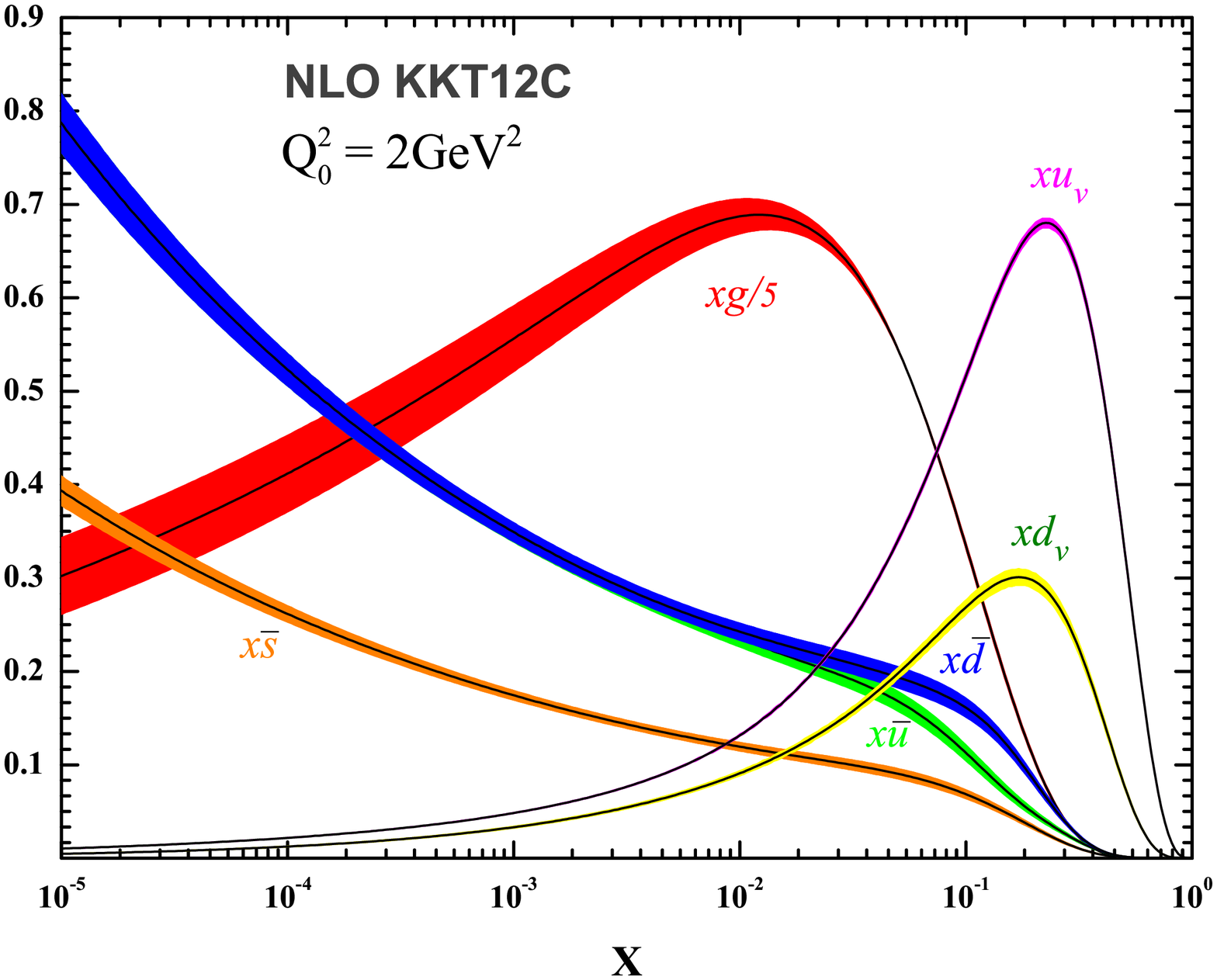}}
\caption{The {\tt KKT12C} parton distributions at input scale
$Q_0^2$ = 2 GeV$^2$ as a function of $x$ in the NLO
approximation.} \label{NLOKKT12C}
\end{figure}

The {\tt KKT12} and {\tt KKT12C} fitted parton distribution
functions at NLO for all sets of parametrizations and their
errors at the starting scale Q$_0^2$ = 2 GeV$^2$ are presented  in
Figs.~\ref{NLOKKT12} and \ref{NLOKKT12C}. The colored bands show
uncertainties on these parton distributions.  By having the
covariance matrix, it is possible to determine the uncertainties
of our PDFs for both fits. Although some of the extracted PDF parameters of two fits are a little
different but in total we found no big differences between the best-fit solutions of the two PDF fits.
But the error bound for some PDFs of two sets are different.

%
%=====================================================================================
%
%
%=====================================================================================
%
%
%=====================================================================================
%

%
%----------------------------------------------------------------
%
\subsubsection{ Comparison with other recent PDF sets }

\begin{figure}
\vspace{1cm}
\centerline{\includegraphics[width=0.48\textwidth]{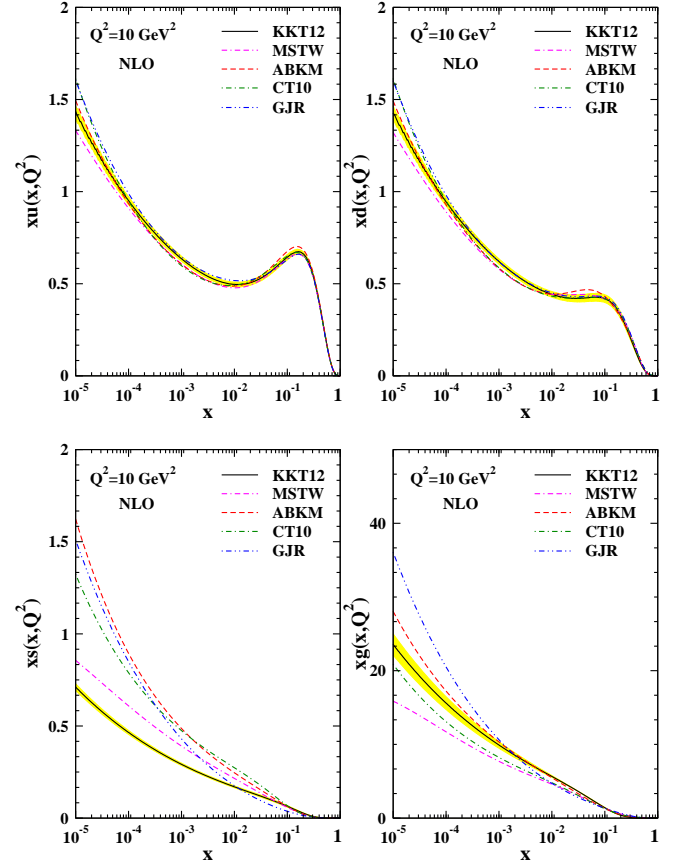}}
\caption{The $u$-, $d$-, $s$-quark and gluon distributions $xg$ at
$Q^2$ = 10 GeV$^2$ as a function of $x$ in the NLO approximation
in comparison to the results obtained by CT10 \cite{Lai:2010vv},
MSTW08 \cite{Martin:2009iq}, ABKM10 \cite{Alekhin:2009ni} and GJR08
\cite{Gluck:2007ck}.}
\label{parton-nlo}
\end{figure}

Our best-fit  {\tt KKT12} PDFs for $u$-, $d$-, $s$-quark and gluon
distributions $xg$ are shown in Figs.~\ref{parton-nlo}  together with a comparison of our results
with CT10 \cite{Lai:2010vv}, MSTW08 \cite{Martin:2009iq}, ABKM10 \cite{Alekhin:2009ni} and GJR08
\cite{Gluck:2007ck}.
%
%=====================================================================================
%
%
%=====================================================================================
%

\begin{figure}
\vspace{1cm}
\centerline{\includegraphics[width=0.48\textwidth]{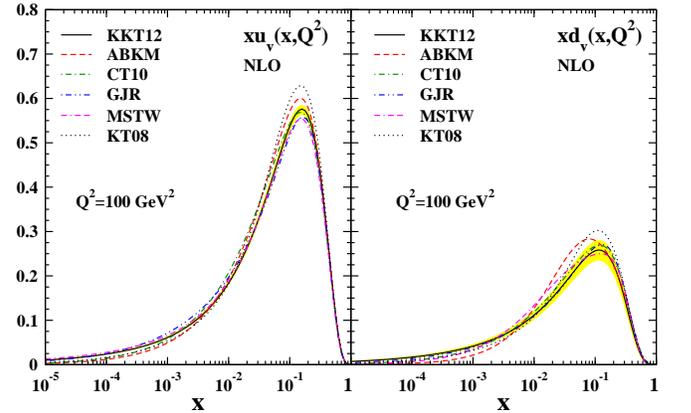}}
\caption{The valence quark distributions $xu_v(x,Q^2)$ and
$xd_v(x,Q^2)$ at $Q^2$ = 100 GeV$^2$ as a function of $x$ in the
NLO approximation which have been compared with the results
obtained by CT10 \cite{Lai:2010vv},
MSTW08 \cite{Martin:2009iq}, ABKM10 \cite{Alekhin:2009ni}, GJR08
\cite{Gluck:2007ck} and valence
analysis of KT08 \cite{Khorramian:2008yh}.}
\label{valance-NLO}
\end{figure}

In Fig.~\ref{valance-NLO} the valence quark distributions
$xu_v(x,Q^2)$ and $xd_v(x,Q^2)$  are shown as a function of $x$ at $Q^2$ =
100 GeV$^2$ in the NLO approximation which have been compared with
the results obtained by CT10 \cite{Lai:2010vv},
MSTW08 \cite{Martin:2009iq}, ABKM10 \cite{Alekhin:2009ni}, GJR08
\cite{Gluck:2007ck} and KT08 \cite{Khorramian:2008yh}.

%
%=====================================================================================
%
%
%=====================================================================================
%

\begin{figure}
\vspace{1cm}
\centerline{\includegraphics[width=0.45\textwidth]{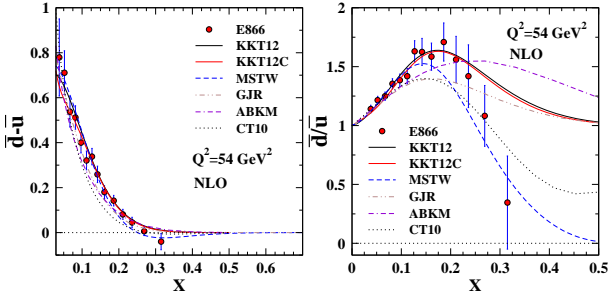}}
\caption{Our NLO results for $\bar{d}-\bar{u}$ and
$\bar{d}/\bar{u}$ as a function of $x$ in comparison to the
results obtained by CT10 \cite{Lai:2010vv},
MSTW08 \cite{Martin:2009iq}, ABKM10 \cite{Alekhin:2009ni} and GJR08
\cite{Gluck:2007ck}. The E866 results
\cite{Towell:2001nh,Webb:2003ps}, scaled to fixed Q$^2$=54
GeV$^2$, are shown as the circles with statistical and systematic
uncertainties.}
\label{dbarubar}
\end{figure}

The NLO results for $\bar{d}-\bar{u}$ and $\bar{d}/\bar{u}$ as a
function of $x$ in comparison to the results obtained by CT10 \cite{Lai:2010vv},
MSTW08 \cite{Martin:2009iq}, ABKM10 \cite{Alekhin:2009ni} and GJR08
\cite{Gluck:2007ck} have been shown in Fig.~\ref{dbarubar}. The E866/NuSea results
\cite{Towell:2001nh,Webb:2003ps}, scaled to fixed Q$^2$=54
GeV$^2$, are shown as the circles with statistical and systematic
uncertainties. The accurate measurements from the E886/NuSea
experiments from $0.02 < x < 0.32$ give clear evidence of the
$\bar{d}-\bar{u}$ asymmetry as seen in Fig.~\ref{dbarubar}. The
asymmetry seems to reach a maximum at $x \approx 0.2$, and becomes
small at smaller $x$ values.

%
%=====================================================================================
%
%
%=====================================================================================
%

%
%----------------------------------------------------------------
%
\subsubsection{ Impact of the H1/ZEUS Combined data }

\begin{figure}
\vspace{1cm}
\centerline{\includegraphics[width=0.40\textwidth]{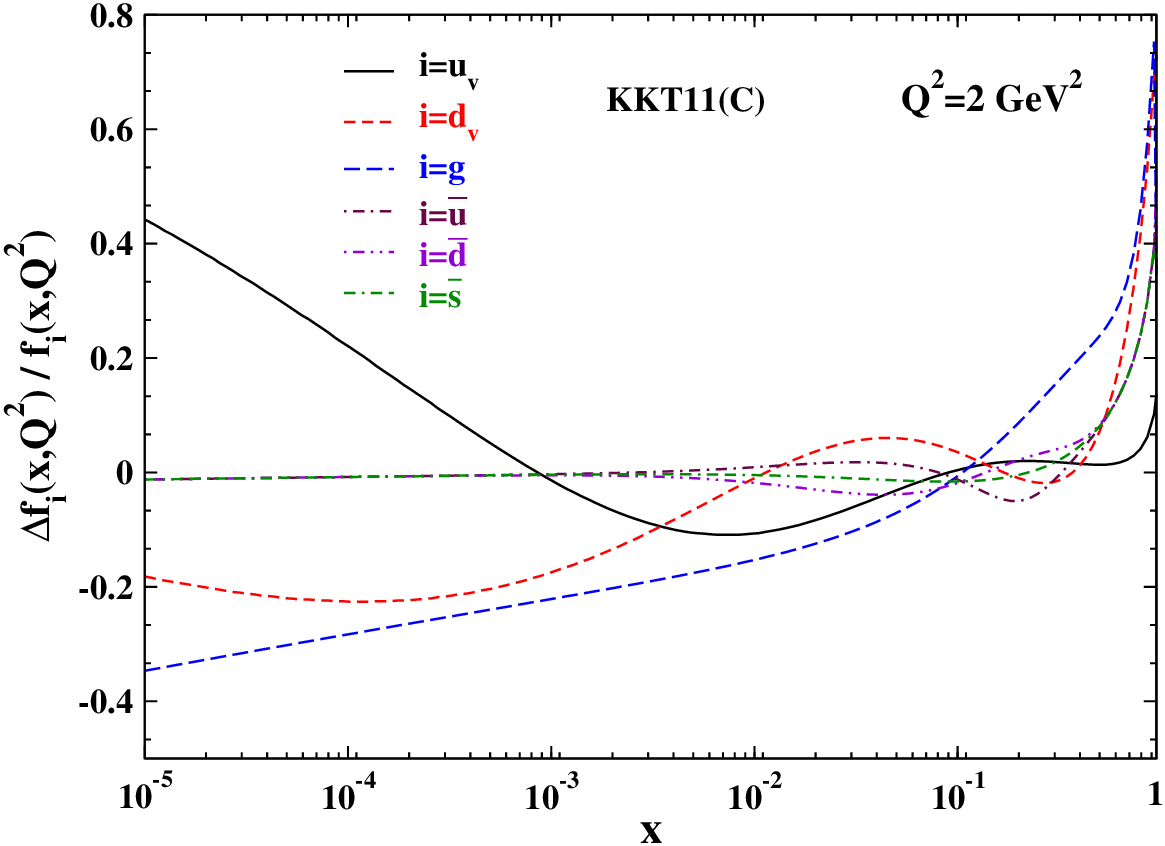}}
\caption{ Ratios of $\Delta f_i(x,Q^2)/f_i(x,Q^2)$ for {\tt KKT11} PDFs (fitted to the separate HERA data sets)
to the {\tt KKT11C} PDFs (fitted to the H1/ZEUS combined data set) at the input scale, $Q_0^2 = 2$ GeV$^2$~\cite{LightCone2011,EPS-HEP2011}.  }
\label{DeltaRatioQ0}
%\end{figure}
%
\vspace{1.7cm}
%=====================================================================================
%
%\begin{figure}
\centerline{\includegraphics[width=0.40\textwidth]{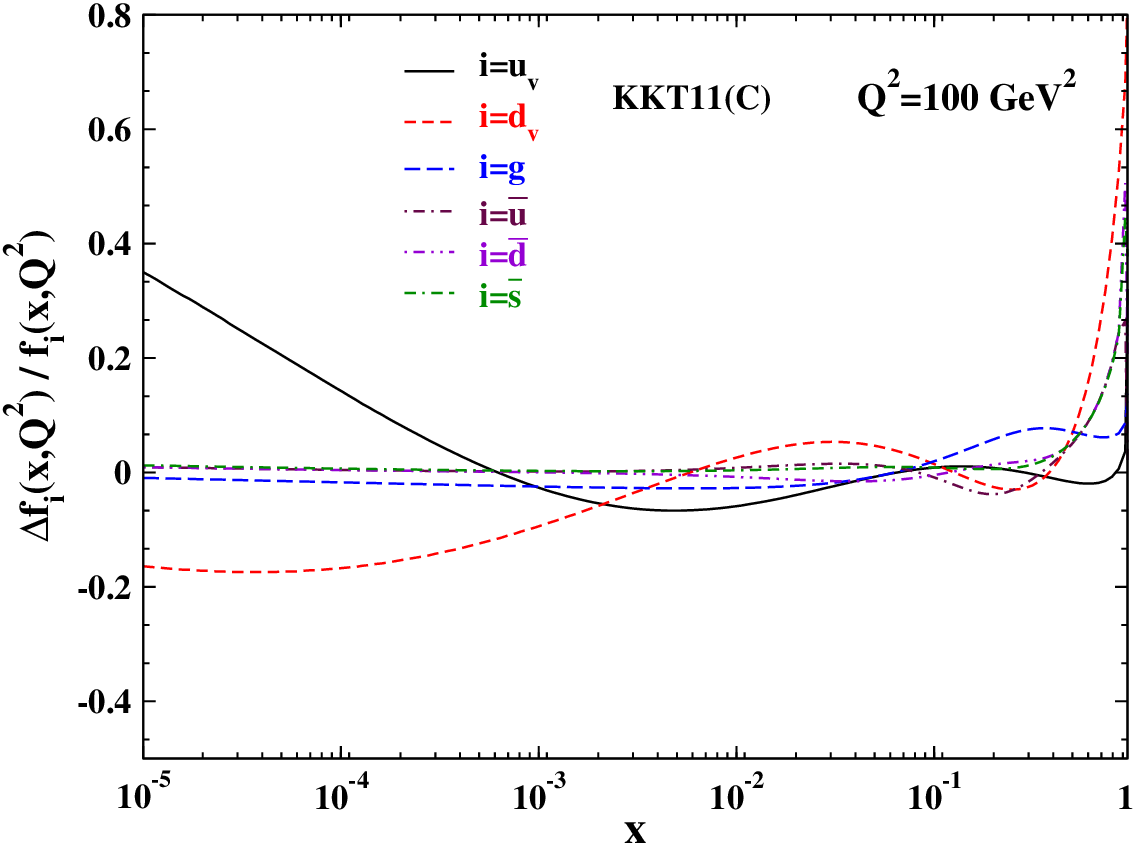}}
\caption{ Ratios of $\Delta f_i(x,Q^2)/f_i(x,Q^2)$ at the higher scale, $Q^2 = 100$ GeV$^2$~\cite{LightCone2011,EPS-HEP2011}. }
\label{DeltaRatioQ100}
\end{figure}

In order to investigate the effect of H1/ZEUS combined $e^{\pm} p$
reduced cross section data on the extracted parton distributions
functions, we show the ratios of the PDFs, $R_i(x,Q^2) =
\Delta f_i(x,Q^2)/f_i(x,Q^2) = \frac{PDF_i^{\tt KKT11} - PDF_i^{\tt KKT11C}}{PDF_i^{\tt KKT11}}$,
in the central PDF sets of the {\tt KKT11C} and {\tt KKT11} fits in
Figs.~\ref{DeltaRatioQ0} and ~\ref{DeltaRatioQ100} at $Q_0^2=2$ GeV$^2$ and
$Q^2=100$ GeV$^2$, respectively. The inclusive jet data from HERA and Tevatron Run-II are not included.
The following results can be found in Refs.~ \cite{LightCone2011,EPS-HEP2011}.
Clearly the effects of the H1/ZEUS combined data sets is on the behavior
of gluon and valence quark PDFs at $x$ below around $10^{-1}$ as the Fig.~\ref{DeltaRatioQ0} shows.
The {\tt KKT11C} gluon distribution at input scale $Q_0^2$ is dominant for small $x$, i.e. $x<0.1$
while the {\tt KKT11} $u$-valance quark PDFs are slightly enhanced at very small
$x$, while the $d$-valance quark PDFs degraded at this region.

In this region, the {\tt KKT11} $\bar{u}$, $\bar{d}$ and $\bar{s}$ distributions are very
close to the corresponding {\tt KKT11C} distributions, therefore $\Delta f_i(x,Q^2)\rightarrow 0$.
In addition the sea-quark ($\bar{u}$-- , $\bar{d}$-- and $\bar{s}$--quarks) PDFs are slightly similar at small
to large $x$ region for these two PDF sets.

Fig.~\ref{DeltaRatioQ100} shows how these ratios are impacted by the DGLAP evolution to higher energy at
$Q^2 = 100$ GeV$^2$. At small to large $x$, the ratios for all quarks and the features of PDFs are similar
to those at $Q_0^2 = 2$ GeV$^2$ described above and have been shown in Figures.
The same behavior of the $u$- and $d$-valance quark PDFs at $Q_0^2 = 2$ GeV$^2$ is repeated for higher scale, $Q^2 = 100$ GeV$^2$.

The differences observed between the two {\tt KKT11C} and {\tt KKT11}
PDFs sets using the H1/ZEUS combined and the separate HERA data
sets are very small, except for the gluon PDFs.
At higher scales we have $\frac {\Delta g(x,Q^2)}{g(x,Q^2)} \rightarrow 0$ as Fig.~\ref{DeltaRatioQ100} shows.

%
%=====================================================================================
%
%
%=====================================================================================
%

\begin{figure}
\vspace{1cm}
\centerline{\includegraphics[width=0.40\textwidth]{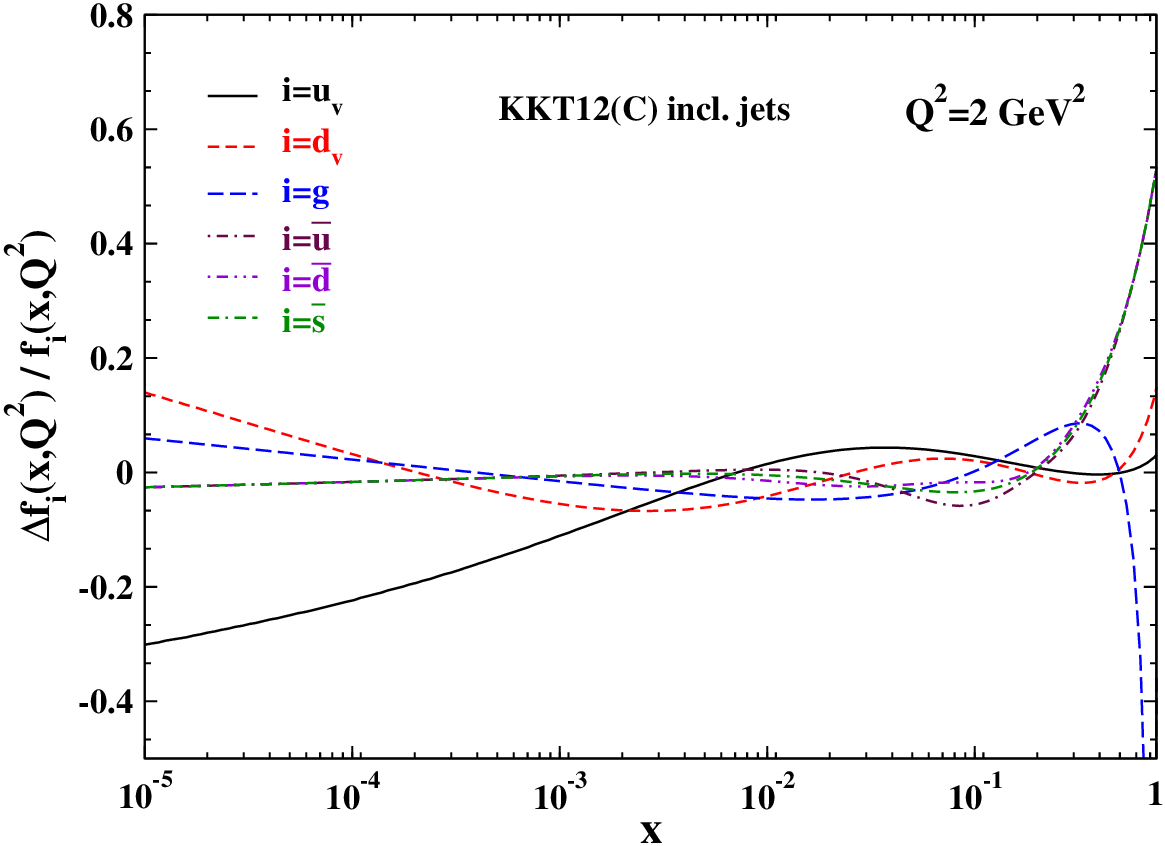}}
\caption{ Ratios of $\Delta f_i(x,Q^2)/f_i(x,Q^2)$ for {\tt KKT12} PDFs
to the {\tt KKT12C} PDFs including jet data at the input scale, $Q_0^2 = 2$ GeV$^2$.  }
\label{DeltaRatioQ0-JET}
%\end{figure}
%
\vspace{1.7cm}
%=====================================================================================
%
%\begin{figure}
\centerline{\includegraphics[width=0.40\textwidth]{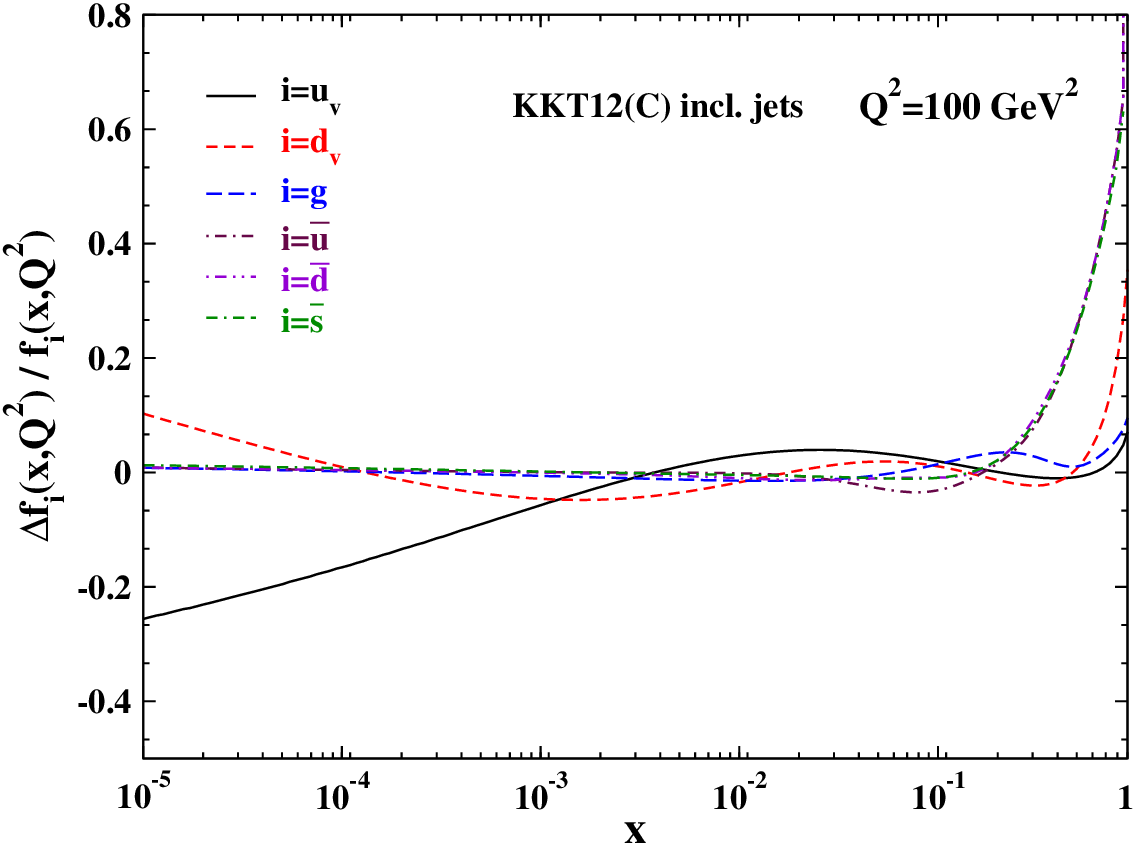}}
\caption{ The same ratios, $\Delta f_i(x,Q^2)/f_i(x,Q^2)$, at the higher scale, $Q^2 = 100$ GeV$^2$. }
\label{DeltaRatioQ100-JET}
\end{figure}

The effect of the H1/ZEUS combined $e^{\pm} p$ reduced cross section data
including the inclusive jet data on extracted PDFs from the present QCD analysis have been shown
in Figs.~\ref{DeltaRatioQ0-JET} and \ref{DeltaRatioQ100-JET}. The results presented in these figures
includes the jet data from HERA~\cite{Chekanov:2002be,Chekanov:2006xr} and
Tevatron~\cite{hep-ex/0011036,Abulencia:2007ez,Abazov:2008hu,hep-ex/0512020,arXiv:0807.2204}.
The differences observed between the results of $\Delta f_i(x,Q^2)/f_i(x,Q^2)$ for {\tt KKT12} PDFs
to the {\tt KKT12C} PDFs at the input $Q_0^2 = 2$ GeV$^2$ and even at higher scale $Q^2 = 100$ GeV$^2$,
for the $\bar{u}$, $\bar{d}$ and $\bar{s}$ distributions are very small and they are slightly similar at small
to large $x$ region.

As Figs.~\ref{DeltaRatioQ0-JET} and \ref{DeltaRatioQ100-JET} show, the mentioned data sets effect the gluon and valence quark PDFs at small $x$, i.e. $x < 10^{-3}$.
The {\tt KKT12} gluon and $d$-valance quark distributions both for input and higher scale are dominant at very small $x$,
while the {\tt KKT12} ratio for $u$-valance quark PDFs are slightly decreases at this region.

%
%=====================================================================================
%
%
%=====================================================================================
%

%\clearpage

%
%%%%%%%%%%%%%%%%%%%%%%%%%%%%%%%%%%%%%%%%%%   Comparison of charm and bottom structure functions to HERA data   %%%%%%%%%%%%%%%%%%%%%%%%%%%%%%%%%%%%%%%%%%%%%%%%%
%
\subsection{ Comparison of charm and bottom structure functions to the HERA data }

\begin{figure}
\vspace{1cm}
\centerline{\includegraphics[width=0.42\textwidth]{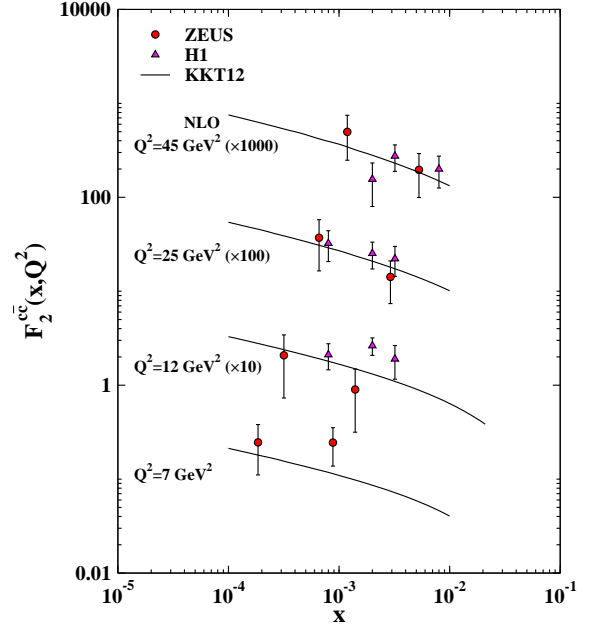}}
\caption{The standard NLO ($\overline{\rm MS}$)
charm quark contribution $F_2^c (x,Q^2)$ in the strict $n_f = 3$ of the FFNS with $m_c =
1.41$. The charm production data are taken from
\cite{Aktas:2004az,Aktas:2005iw,Adloff:2001zj,Webb:2003ps}. }
\label{F2charm1}
\end{figure}

\begin{figure}
\vspace{1cm}
\centerline{\includegraphics[width=0.35\textwidth]{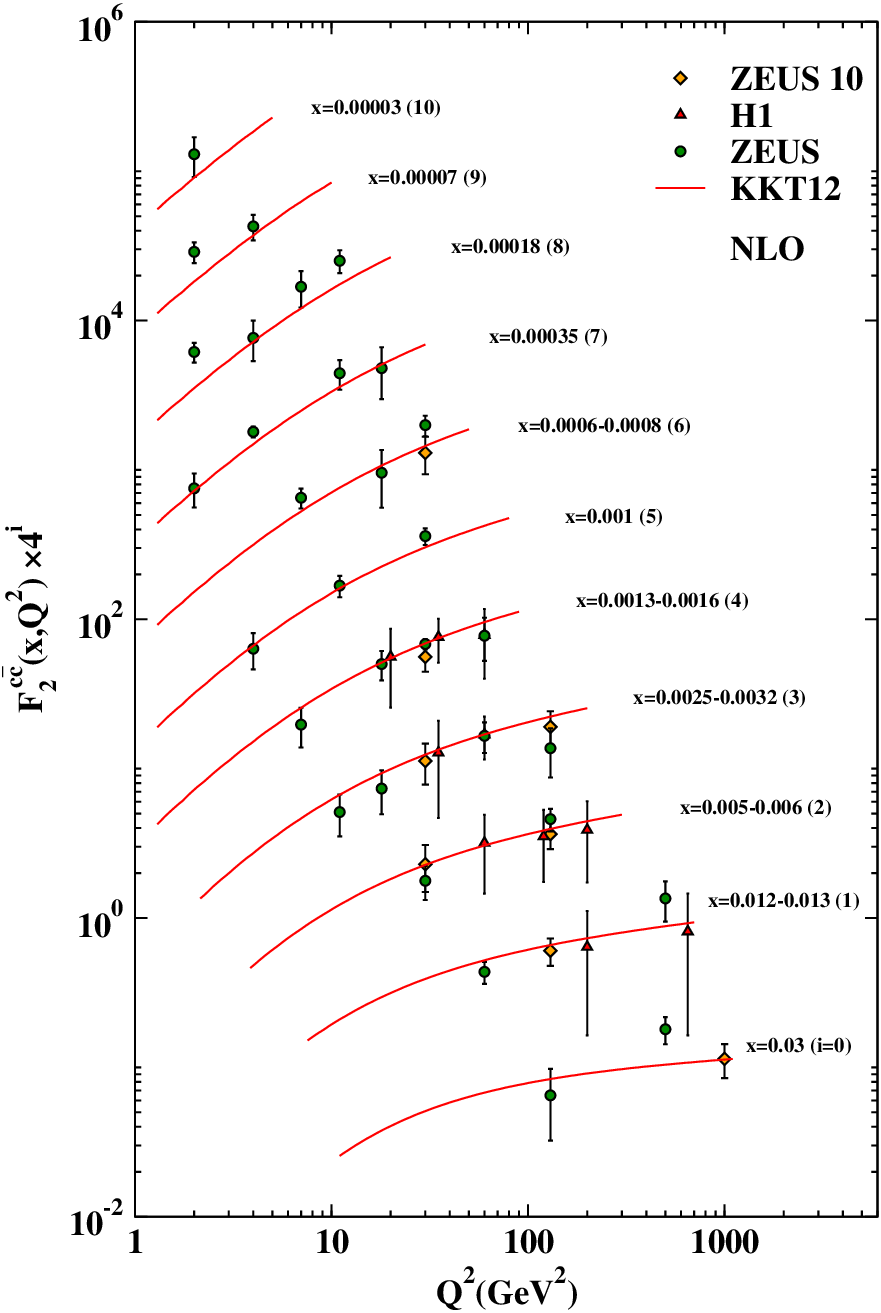}}
\caption{\small The standard NLO ($\overline{\rm MS}$) charm
contribution $F_2^c (x,Q^2)$ to the proton structure function
shown as a function of Q$^2$ for various $x$ value.
For comparison
we also display the charm production data
\cite{Aktas:2004az,:2009ut,Aktas:2005iw,Adloff:2001zj,Adloff:1996xq,Chekanov:2007ch,Breitweg:1999ad,Chekanov:2003rb}.
}
\label{F2charm2}
\end{figure}

Charm and bottom contributions to the proton structure functions, $F_2^c (x,Q^2)$ and $F_2^b (x,Q^2)$, have been
extracted from Eq.~\ref{eq:FFNS}. The detailed discussion of corresponding contributions
have been presented in Sec.~\ref{Heavy-flavour-contributions}.
The NLO ($\overline{\rm MS}$) charm contribution $F_2^c (x,Q^2)$
to the proton structure function in the strict $n_f = 3$ of the
FFNS with $m_c = 1.41$ as a function $x$ for different
values of Q$^2$ and as a function of Q$^2$
for various $x$ values are shown in Figs.~\ref{F2charm1} and \ref{F2charm2}, respectively. For comparison we
also display the charm production data
\cite{Aktas:2004az,:2009ut,Aktas:2005iw,Adloff:2001zj,Adloff:1996xq,Chekanov:2007ch,Breitweg:1999ad,Chekanov:2003rb}
as well.

%
%=====================================================================================
%

%
%=====================================================================================
%

%
%=====================================================================================
%

\begin{figure}
\vspace{1cm}
\centerline{\includegraphics[width=0.47\textwidth]{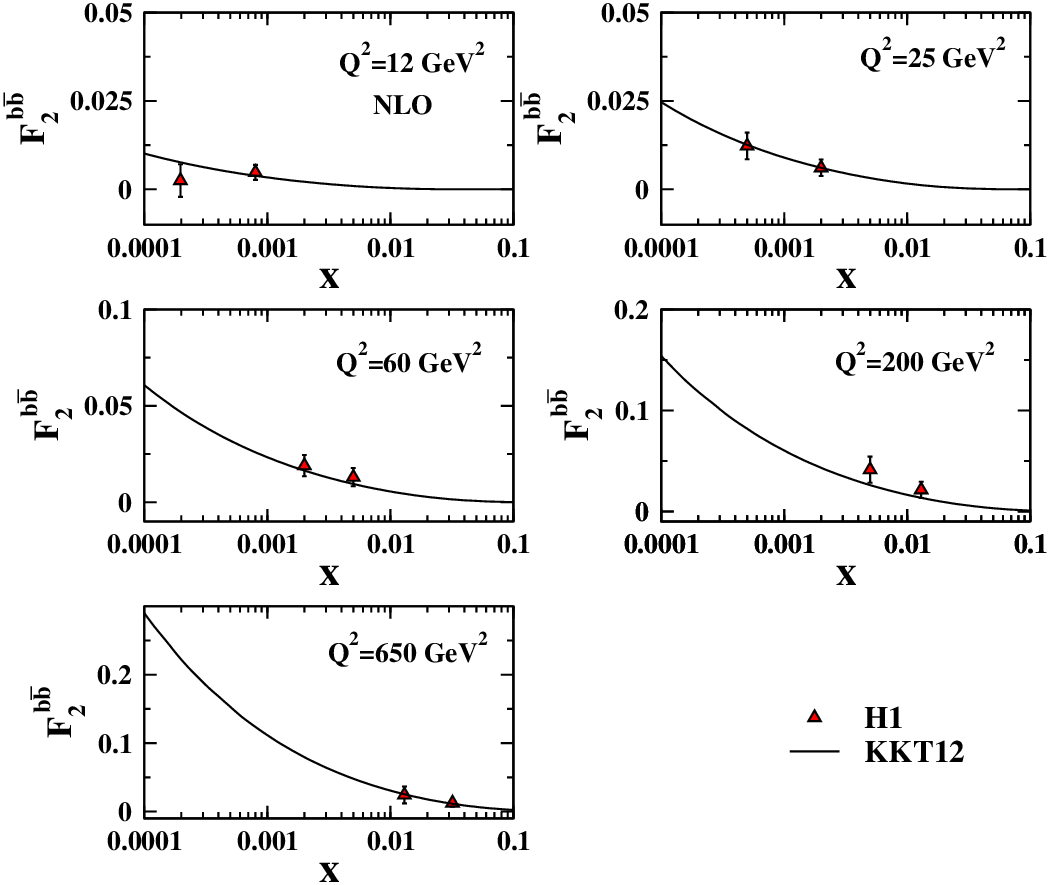}}
\caption{The standard NLO ($\overline{\rm MS}$) bottom
contribution $F_2^b (x,Q^2)$ to the proton structure function
shown as a function of Q$^2$ for various $x$ value. For comparison
we also display the charm production data
\cite{Aktas:2004az,Aktas:2005iw}. }
\label{F2bottom}
\end{figure}

\begin{figure}
\vspace{1.2cm}
\centerline{\includegraphics[width=0.45\textwidth]{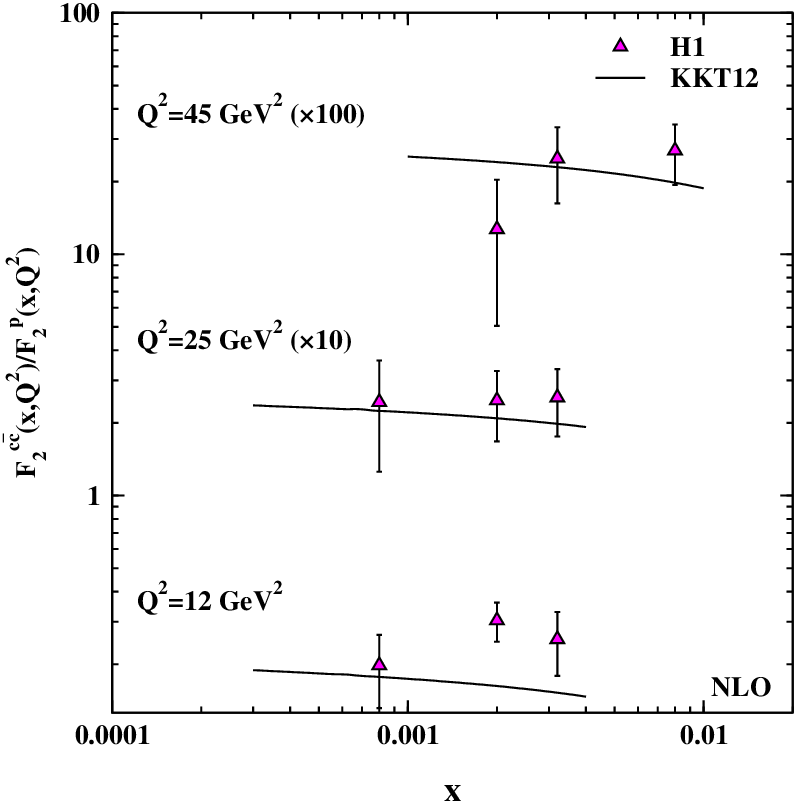}}
\caption{The structure function ratio
$\frac{F_2^c(x,Q^2)}{F_2^p(x,Q^2)}$ as a function of $x$ at Q$^2$ =
12, 25, 45 Ge$V^2$ in comparison with H1 data
\cite{Adloff:1996xq}. }
\label{f2ccoverf2p}
\end{figure}

Fig.~\ref{F2bottom} shows the NLO ($\overline{\rm MS}$) bottom contribution $F_2^b (x,Q^2)$ to the
proton structure function shown as a function of $x$ for various
Q$^2$ value in comparison with the charm production data
\cite{Aktas:2004az,Aktas:2005iw}. The structure function ratio
$\frac{F_2^c(x,Q^2)}{F_2^p(x,Q^2)}$ as a function of $x$ at Q$^2$ =
12, 25, 45 Ge$V^2$ in comparison with H1 data \cite{Adloff:1996xq}
has been shown in Fig.~\ref{f2ccoverf2p}.

%
%=====================================================================================
%

%
%=====================================================================================
%

%
%=====================================================================================
%

%
%%%%%%%%%%%%%%%%%%%%%%%%%%%%%%%%%%%%%%%%%%   Comparison to the inclusive-jet and dijet data   %%%%%%%%%%%%%%%%%%%%%%%%%%%%%%%%%%%%%%%%%%%%%%%%%
%
\subsection{ Comparison to the inclusive jet data }

\begin{figure}
\vspace{1cm}
\centerline{\includegraphics[width=0.44\textwidth]{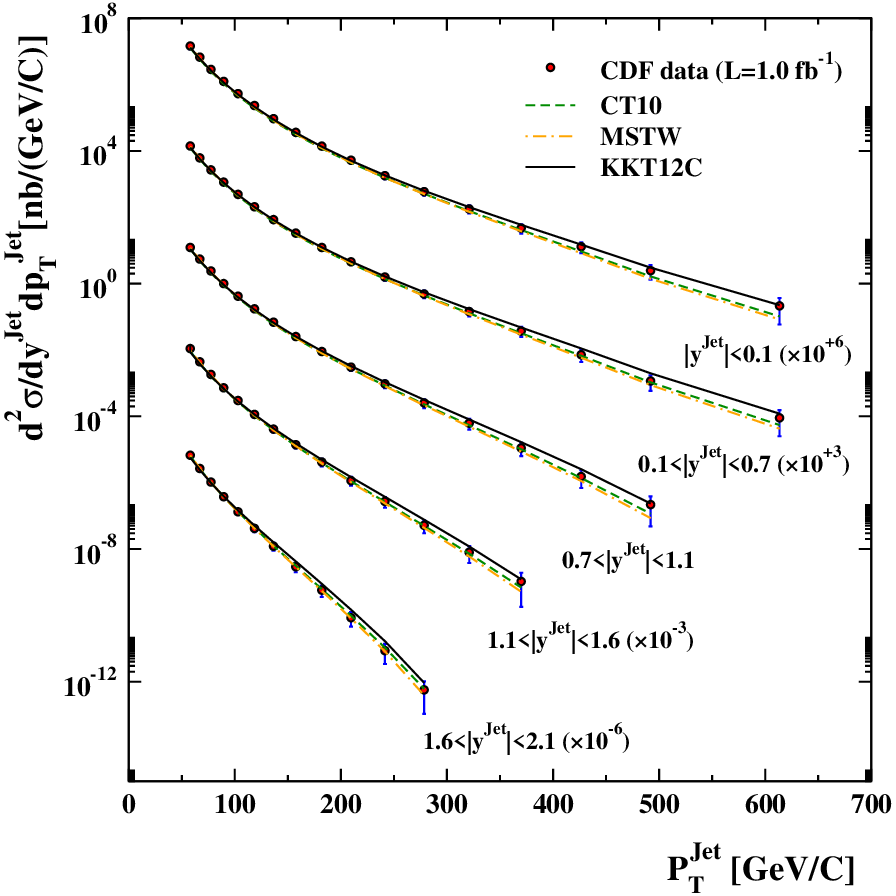}}
\caption{ The CDF inclusive jet cross sections measured (circles) as a function of $p_{T}^{jet}$
for jets with $p_{T}^{jet} > 54$~GeV/c
in different $|y^{jet}|$ regions compared to NLO pQCD predictions based on the {\tt KKT12C} PDF (black lines). }
\label{JETCDF}
\end{figure}

\begin{figure}
\vspace{1cm}
\centerline{\includegraphics[width=0.44\textwidth]{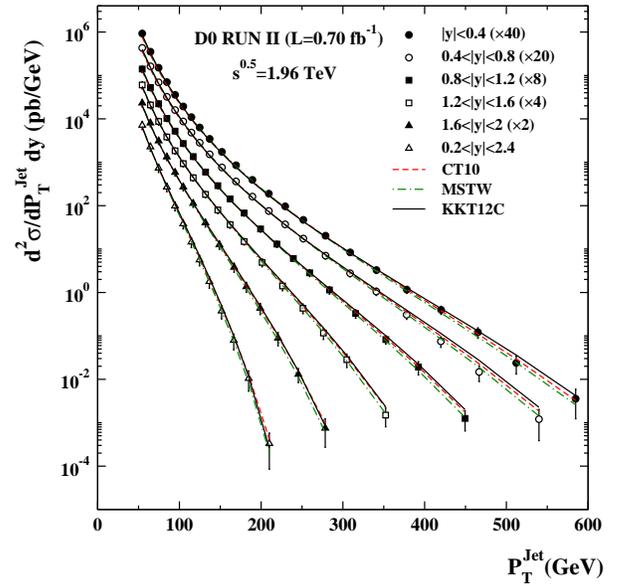}}
\caption{ The D{\O} inclusive jet cross sections in six rapidity regions compared to our
NLO pQCD predictions based on the {\tt KKT12C} PDF. The data points are multiplied by 40, 20, 8, 4, and 2
for the bins $|y|<0.4$, $0.4<|y|<0.8$, $0.8<|y|<1.2$, $1.2<|y|<1.6$, and  $1.6<|y|<2.0$, respectively.}
\label{JETD0}
\end{figure}

As indicated in Sec.~\ref{data-sets}, the high-$x$ gluon distribution in our global QCD fits is determined by the
HERA and Tevatron Run-II inclusive jet data. These data sets have the potential to constrain the gluon
density in the proton when used as inputs to global fits of the proton parton distribution functions.
In order to illustrate the precision of the {\tt KKT12C} PDFs, we compare the total cross
sections of some selected  jet production processes at the Tevatron Run-II from CDF and D{\O}.
Our {\tt KKT12C} NLO QCD predictions for inclusive jet cross sections are compared to the Tevatron Run-II measurements.
Fig.~\ref{JETCDF} shows the CDF measured inclusive jet cross sections~\cite{Abulencia:2007ez}, $d^{2}\sigma /dp_{T}^{jet}dy^{jet}$, as a
function of $p_{T}^{jet}$ in five different $|y^{jet}|$ regions compared to our {\tt KKT12C} NLO pQCD predictions.
For presentation, the measurements in different $|y^{jet}|$ regions are scaled by different
global factors as indicated in parentheses in the figure.
Overall the {\tt KKT12C} predictions of NLO QCD gives a good description of the data.
The D{\O} Collaboration also reported their measurement
of the inclusive jet production cross section~\cite{Abazov:2008hu}, in which
a comparison was made to our NLO theory calculation (with {\tt fastNLO} code ~\cite{Kluge:2006xs}) based on {\tt NLOJET++}~\cite{Nagy:2001fj,Nagy:2003tz}
using the {\tt KKT12C} PDF set.
The corresponding results of the inclusive jet cross section measurement are displayed in Fig.~\ref{JETD0}
in six $|y^{jet}|$ bins as a function of $p_{T}^{jet}$.
The comparison shows a good agreement between our NLO theoretical predictions and the data.

%
%=====================================================================================
%
%
%=====================================================================================
%

%
%=====================================================================================
%
%
%=================================================================%====================
%

%
%
%%%%%%%%%%%%%%%%%%%%%%%%%%%%%%%%%%%%%%%%%%    The longitudinal structure function    %%%%%%%%%%%%%%%%%%%%%%%%%%%%%%%%%%%%%%%%%%%%%%%%%
%
\section{The longitudinal structure function $F_L(x,Q^2)$ } \label{longitudinalSF}
In this section we turn to the perturbative predictions for longitudinal structure function
$F_L(x,Q^2)$. Similarly to $F_2^p(x,Q^2)$ one can write for $F_L(x,Q^2)$ in the
common $\overline{\rm MS}$ factorization scheme for $n_f = 3$ light quark flavour,

\begin{eqnarray}
x^{-1}F_L(x,Q^2) = C_{L,{\rm NS}} \otimes  \Big(\frac{1}{6}\ q_{3}^+ +\frac{1}{18}\, q_{8}^+\Big) \hspace{1.5cm} \nonumber \\
+ \frac{2}{9} \left( C_{L,q}\otimes\Sigma +  C_{L,g}\otimes g\right) + x^{-1}F_L^{\rm heavy}(x,Q^2)\,. \, \,
\end{eqnarray}

Since the longitudinal structure functions $F_L(x,Q^2)$ contains rather large heavy flavour contributions
in the small-$x$ region, a consistent analysis has to be taken into account for these effects.
The total longitudinal structure function can be written as a sum of $F_L(x,Q^2) = F_L^{light}(x,Q^2) + F_L^{heavy}(x,Q^2)$ where
heavy quark contributions are $F_L^{heavy} = F_L^c(x,Q^2) + F_L^b(x,Q^2)$, and the top quark contribution is negligible.

%
%=====================================================================================
%
\begin{figure}[tbh]
\vspace{1.0cm}
\centerline{\includegraphics[width=0.42\textwidth]{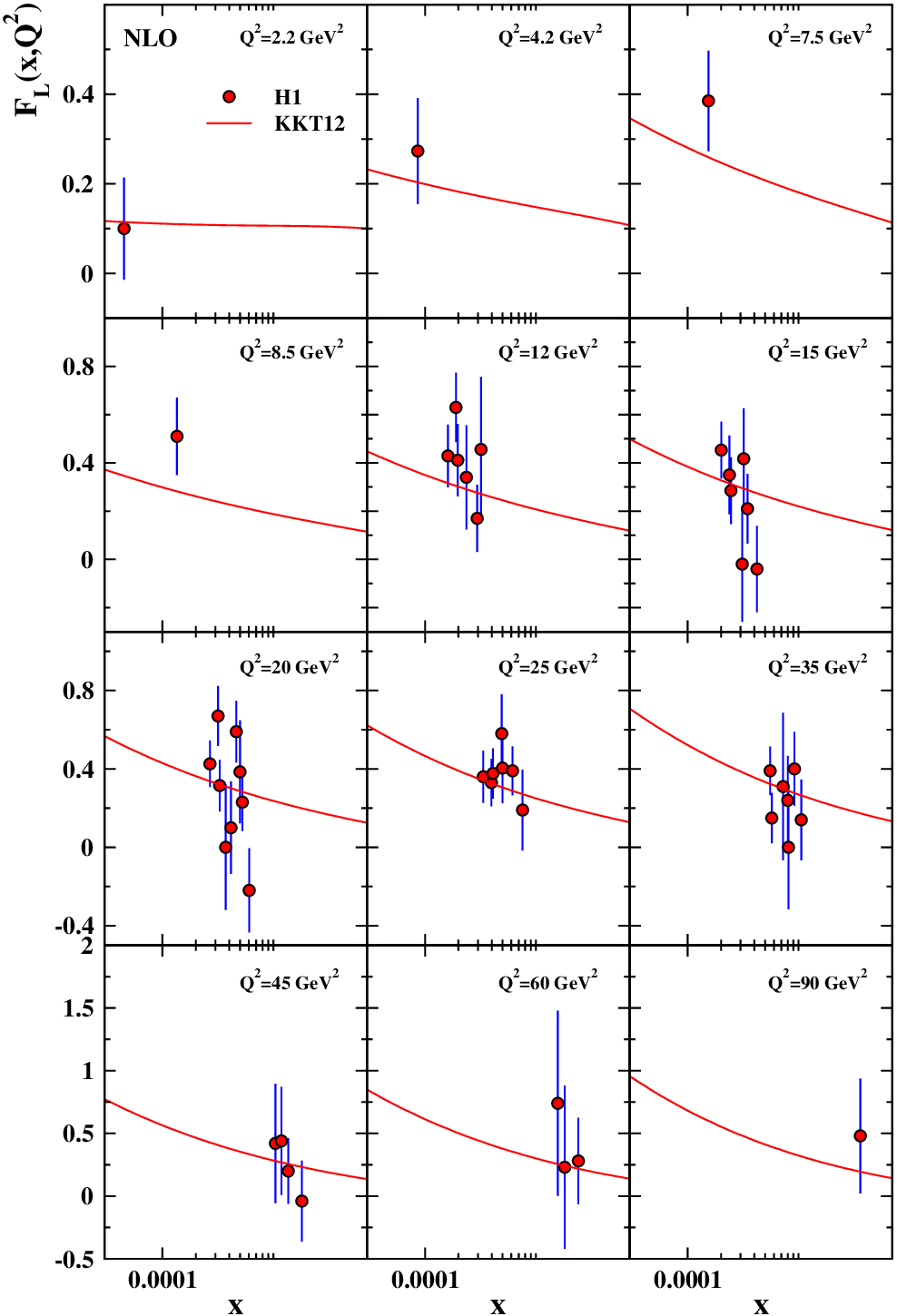}}
\caption{Predictions for the longitudinal proton structure
function, $F_L(x,Q^2)$, at NLO in comparison with H1 data
\cite{Adloff:2000qk,Adloff:2003uh,Adloff:1996yz,:2008tx}.}
\label{FLH1}
\end{figure}
%
%=====================================================================================
%

The extracted parton distributions from present analysis have been
used to predict $F_L(x,Q^2)$. The prediction for the longitudinal structure functions $F_L(x,Q^2)$ is mainly determined by the form of the gluon
distribution extracted from the global QCD analysis. Measuring the structure
function $F_L$ therefore provides a way of studying the gluon
density and a test of perturbative QCD. The study of $F_L(x,Q^2)$,
led us to extract heavy quarks, especially gluon distribution
function at low $x$.

\begin{figure}[tbh]
\vspace{1.0cm}
\centerline{\includegraphics[width=0.45\textwidth]{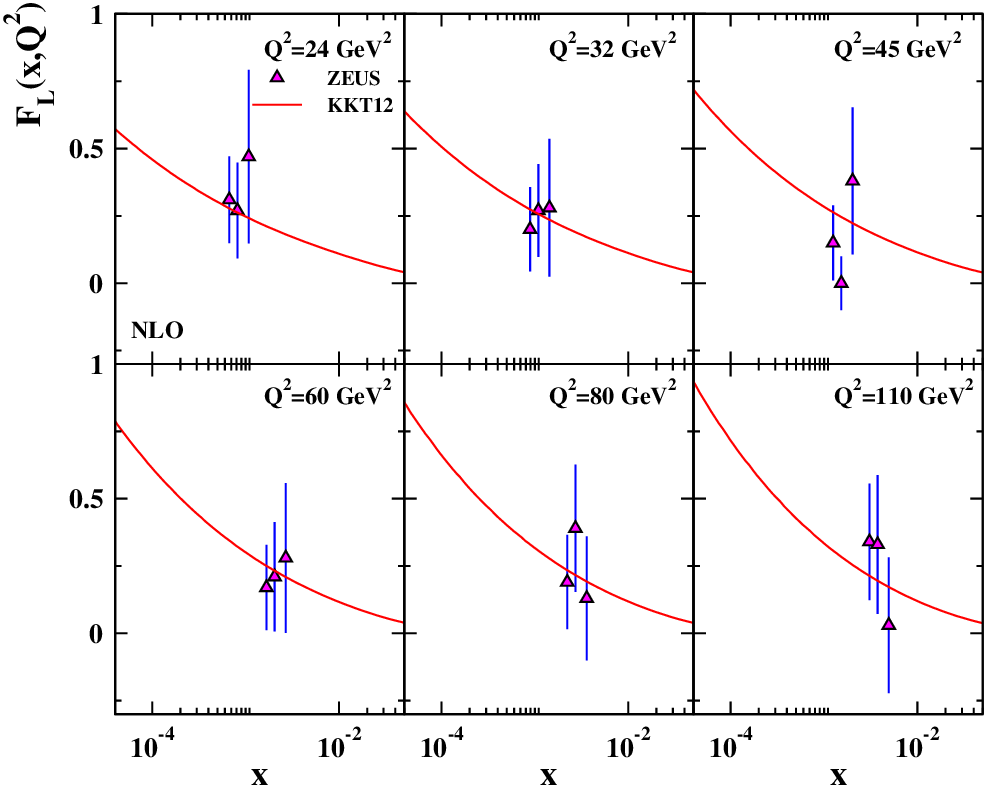}}
\caption{The longitudinal proton structure
function, $F_L(x,Q^2)$, at NLO in comparison with ZEUS data
\cite{Chekanov:2009na}.}
\label{FLZEUS}
\end{figure}

For completeness, we finally shown our NLO QCD standard
predictions for $F_L(x,Q^2)$ with a representative selection of
HERA-H1 data \cite{Adloff:2000qk,Adloff:2003uh,:2008tx,Adloff:1996yz} in
Fig.~\ref{FLH1} and HERA-ZEUS data \cite{Chekanov:2009na} in
Fig.~\ref{FLZEUS} which we used in our QCD analysis.

%
%=====================================================================================
%
%
%=====================================================================================
%

\begin{figure}[tbh]
\vspace{1.0cm}
\centerline{\includegraphics[width=0.45\textwidth]{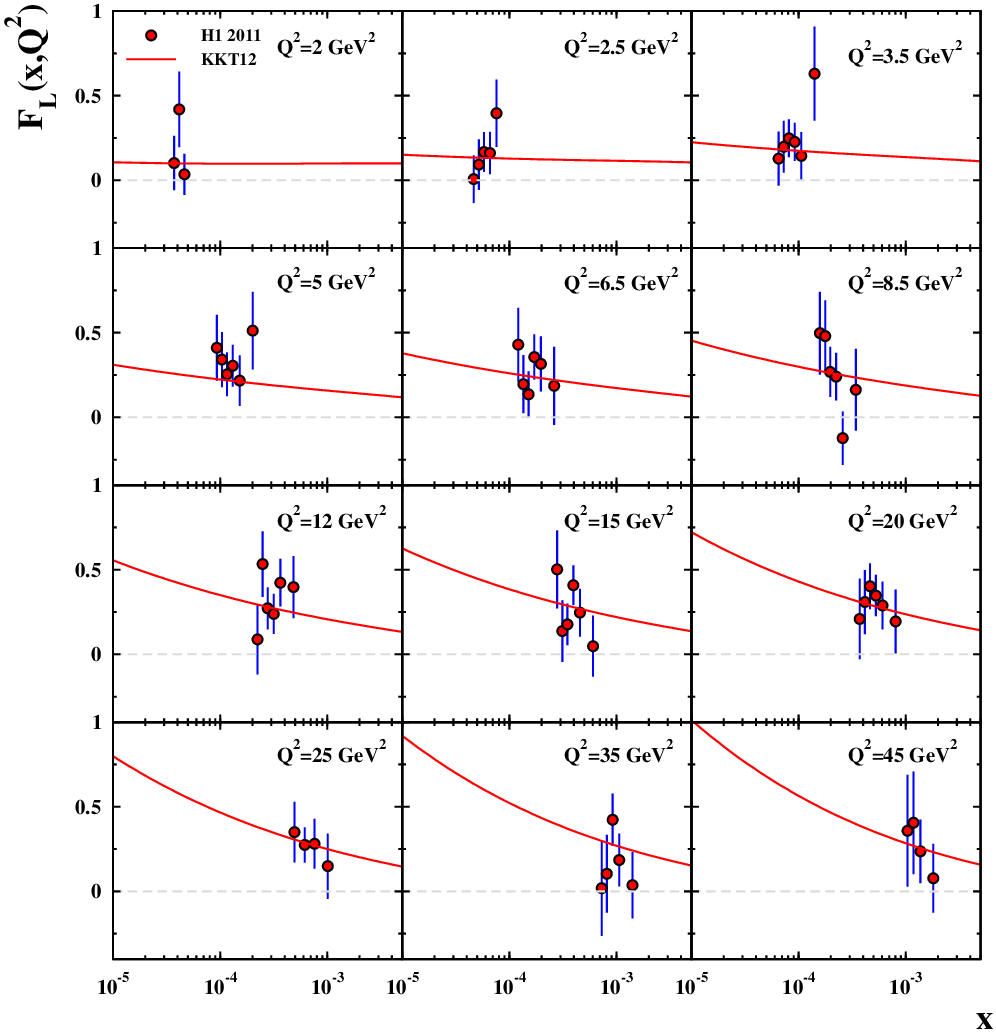}}
\caption{Predictions for the longitudinal proton structure
function, $F_L(x,Q^2)$, at NLO in comparison with recent H1 data
\cite{Collaboration:2010ry}.}
\label{FL2011}
\end{figure}

The predicted longitudinal structure function $F_L(x,Q^2)$ as a function of $x$  has
been also compared with the recent H1 data
\cite{Collaboration:2010ry} in Fig.~\ref{FL2011}. Our results for
$F_L(x,Q^2)$, being gluon dominated in the small-$x$ region, are
in full agreement with present measurements of $F_L(x,Q^2)$.

%
%=====================================================================================
%
%
%=====================================================================================
%

\begin{figure}
\vspace{1.0cm}
\centerline{\includegraphics[width=0.43\textwidth]{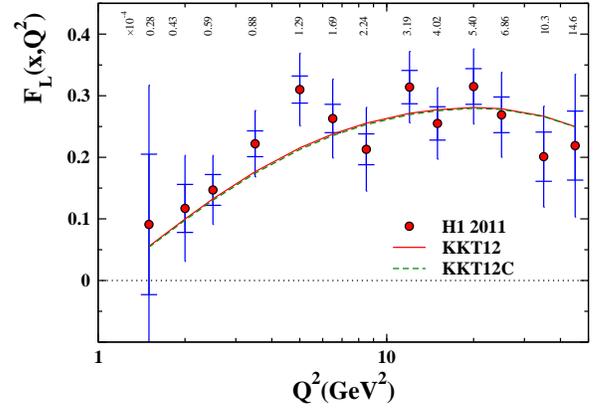}}
\caption{The longitudinal structure function $F_L(x,Q^2)$ shown as a function of $Q^2$ for
different values of averaged $x$ has been compared to the most recent H1 data for
longitudinal proton structure function \cite{Collaboration:2010ry}.
The full error bars in the figure include the statistical
and systematic uncertainties added in quadrature.
The inner error bars only represent statistical error. }
\label{FL-H1-2011}
\end{figure}

The longitudinal structure function $F_L(x,Q^2)$ as a function of $Q^2$ and
for different values of averaged $x$ are shown with the most
recent H1 \cite{Collaboration:2010ry} data in Fig.~\ref{FL-H1-2011}. The full error bars in this figure include
the statistical and systematic uncertainties added in quadrature.
The inner error bars represent statistical error.

%
%=====================================================================================
%
%
%=====================================================================================
%

\clearpage
%
%
%%%%%%%%%%%%%%%%%%%%%%%%%%%%%%%%%%%%%%%%%%    Determination of the Strong Coupling    %%%%%%%%%%%%%%%%%%%%%%%%%%%%%%%%%%%%%%%%%%%%%%%%%
%
\section {Determination of the strong coupling constant}  \label{strong-coupling}

The determination of $\alpha_s$ is still an open problem in PDF fits
since it's treatment is closely related to the most important
processes at hadron colliders such as the LHC. This importance is
due to the dependance of such processes and their partonic cross
section to strong coupling constant,  $\alpha_s$. We include $\alpha_s({\rm Q_0^2})$ as a free
parameter in our fits, thus the running coupling constant is
determined in our analysis together with the parton distributions
of the nucleon; in particular it is closely related to the gluon
distribution which drives the QCD evolution. Although some theoretical
groups, like CT10 \cite{Lai:2010vv} or NNPDF \cite{Ball:2012cx,Ball:2010de}, fix the $\alpha_s$ value close to
the updated Particle Data Group (PDG) average as $\alpha_s({\rm M_Z^2}) = 0.1184 \pm 0.0007$  \cite{Nakamura:2010zzi}, but other theoretical
groups such as MSTW08 \cite{Martin:2009iq}, ABKM10 \cite{Alekhin:2009ni} or
GJR08 \cite{Gluck:2007ck} determine $\alpha_s$  as a free parameter in the QCD fits.

The values for the $\alpha_s({\rm M_Z^2})$ are obtained in {\tt KKT12} fit are slightly larger than {\tt KKT12C} QCD analysis.
This is due to different data sets which we used in our fits.
The gluon distributions extracted with and without
HERA combined data sets play an important role in $\alpha_s(\rm M_Z^2)$ determination.
In addition, inclusion of the inclusive jet data also effected our $\alpha_s({\rm M_Z^2})$ values.
In our previous QCD analysis based on separate and combined data sets, which we didn't include
the HERA and Tevatron jet data, we obtained $\alpha_{s}(M_{Z}^{2}) = 0.1185 \pm 0.0021$
for {\tt KKT11} to be compared with $\alpha_{s}(M_{Z}^{2}) = 0.1165 \pm
0.0013$ for {\tt KKT11C} at the NLO~\cite{LightCone2011,EPS-HEP2011}.

Comparison of the present $\alpha_s(\rm M_Z^2)$ determination at NLO with the
values obtained by other PDF fitting groups has been shown in the
Table~\ref{Table3}. The $\alpha_s({\rm M_Z^2})$ values of Ref.
\cite{Martin:2009iq}  are slightly larger than the corresponding
results of both sets of our fits. The {\tt KKT12} and {\tt KKT12C}
values of $\alpha_s({\rm M_Z^2})$ are both consistent with the
world average value of $\alpha_s({\rm M_Z^2}) = 0.1176 \pm 0.0020$ \cite{Amsler:2008zzb}.

%
%=====================================================================================
%
\begin{table}[tbh]
\centering
{\begin{tabular}{@{}c|lccl@{}}
 \hline \hline
%--------------------------------------------------------------------------------
     &                  & Reference  &  $\alpha_s({\rm M_Z^2})$        & Notes                        \\        \hline
     &  KKT12           &            &  $0.1149 \pm 0.0019 $           & standard~approach            \\
     &  KKT12C          &            &  $0.1142 \pm 0.0027 $           & standard~approach            \\
     & MSTW08           & \cite{Martin:2009iq}  &
$0.1202 {\footnotesize{\begin{array}{c} +0.0012                 \\
-0.0015 \end{array}}}$    &                                    \\
         & KT08         & \cite{Khorramian:2008yh} &  $0.1149 \pm 0.0021$   & valence~analysis        \\
         & GJR08        & \cite{Gluck:1998xa}      &  $0.1178 \pm 0.0021$   & standard~approach       \\
NLO  & H1           & \cite{Adloff:2000qk}     &  $0.1150 \pm 0.0017$   &                         \\
         & CTEQ         & \cite{Pumplin:2005rh}    &  $0.1170 \pm 0.0047$   &                         \\
         & Alekhin      & \cite{Alekhin:2002fv}    &  $0.1171 \pm 0.0015$   &                         \\
         & ZEUS         & \cite{Chekanov:2005nn}   &  $0.1183 \pm 0.0028$   &                         \\
         & BBG          & \cite{Blumlein:2006be}   &  $
0.1148 {\footnotesize{\begin{array}{c} +0.0019                 \\
-0.0019 \end{array}}}$   &  valence~analysis                   \\         \hline  \hline
%--------------------------------------------------------------------------------
\end{tabular}}
\caption{Comparison of the present $\alpha_s(\rm M_Z^2)$ determination at NLO with the values obtained by other PDF fitting groups.}
\label{Table3}
\end{table}
%
%=====================================================================================
%

%
%%%%%%%%%%%%%%%%%%%%%%%%%%%%%%%%%%%%%%%%%%    Summary and conclusions    %%%%%%%%%%%%%%%%%%%%%%%%%%%%%%%%%%%%%%%%%%%%%%%%%
%
\section{Summary and conclusions} \label{Summary}
In the present paper, the most up-to-date data on deep
inelastic scattering and related processes for the structure
functions $F_{2}^{i}(x,Q^2)$ with $(i=p, d, c, b)$, $F_{L}
(x,Q^2)$, $xF_{3}(x,Q^2)$ and `reduced' cross-section data
have been analyzed in the standard NLO parton model approach of the perturbative QCD.
We have produced two new PDF sets, {\tt KKT12} and {\tt KKT12C} which include up-to-date
DIS data as well as the HERA and Tevatron Run-II inclusive jet cross section data.
The {\tt KKT12C} is obtained using the HERA combined data sets while the
{\tt KKT12} included the HERA results as separate data sets.

Our predictions for all PDFs using QCD fits to the data from all available experimental data
and the most recent HERA combined measurements for {\tt KKT12} and {\tt KKT12C}
respectively, are in very good agreement with the other theoretical
models. The strong coupling constant obtained from our standard
NLO analysis in the FFN scheme with $n_f = 3$ active light ($u$,
$d$, $s$) flavours is $\alpha_{s}(M_{Z}^{2}) = 0.1149 \pm 0.0019$
({\tt KKT12}) to be compared with $\alpha_{s}(M_{Z}^{2}) = 0.1142 \pm
0.0027$ ({\tt KKT12C}).

A FORTRAN package (grid) containing both our sets of standard NLO
($\overline{\rm MS}$) parton densities, the light $F_2^{light} (x,Q^2)$ as well as the heavy $F_2^{heavy}
(x,Q^2)$  and longitudinal $F_L(x,Q^2)$ and $xF_3$ structure
functions can be obtained
by electronic mail or can be found directly from
{\tt http://particles.ipm.ir/links/QCD.htm} \cite{Program-summary}.
This FORTRAN package covers the following ranges in $x$ and Q$^2$,
10$^{-9}$ $\leq$ x $\leq$ 1 and 1 GeV$^2$ $\leq$ Q$^2$ $\leq$
10$^6$ GeV$^2$ define for high $Q^2$ and/or low $x$.
The QCD analysis based on the NNLO approximation is in progress.

%\clearpage

%
%%%%%%%%%%%%%%%%%%%%%%%%%%%%%%%%%%%%%%%%%%    Acknowledgments    %%%%%%%%%%%%%%%%%%%%%%%%%%%%%%%%%%%%%%%%%%%%%%%%%
%
\section*{Acknowledgments}
We are especially grateful to F. Olness for valuable discussion and critical remarks.
A.N.K thanks the CERN TH-PH division for the hospitality
where a portion of this work was performed. We acknowledge
financial support of Semnan university and also the School of
Particles and Accelerators, Institute for Research in Fundamental
Sciences (IPM).
%
%

%
%%%%%%%%%%%%%%%%%%%%%%%%%%%%%%%%%%%%%%%%%%    Bibliography    %%%%%%%%%%%%%%%%%%%%%%%%%%%%%%%%%%%%%%%%%%%%%%%%%
%
%

\end{document}